\documentclass[12pt, a4paper]{article}
\usepackage{amsmath}
\usepackage{amssymb}
\usepackage{libertine}
\usepackage{titling}
\usepackage{graphicx}
\usepackage{mathrsfs}
\usepackage{tikz}
\usepackage{pgfplots}
\usepackage{bm}
\usepackage{subcaption}
\usepackage{adjustbox}
\usepackage{threeparttable}
\usepackage{listings}
\usepackage{amsthm}
\usepackage{accents}
\usepackage{soul}
\usepackage{mathtools,xparse}
\usepackage[a4paper, top=1.cm, bottom=1.cm, left=1.8cm, right=1.8cm, footskip = 1cm, includehead, includefoot]{geometry}
\usepackage{lipsum}
\usepackage[parfill]{parskip} 
\usepackage{centernot}
\usepackage{color}
\usepackage[T1]{fontenc}
\usepackage[scaled=.85]{beramono}
\usepackage[toc, title, titletoc]{appendix}

\makeatletter
\renewcommand{\@makefntext}[1]{%
  \setlength{\parindent}{0pt}%
  \begin{list}{}{%
    \setlength{\labelwidth}{.5em}
    \setlength{\leftmargin}{0.055\textwidth}%
    \setlength{\labelsep}{2pt}%
    \setlength{\itemsep}{0pt}%
    \setlength{\parsep}{0pt}%
    \setlength{\topsep}{0pt}%
  \setlength{\rightmargin}{0.075\textwidth}%
    \footnotesize}%
  \item[\@makefnmark\hfil]#1%
  \end{list}%
}
\makeatother

\usepackage[colorlinks=true,
            linkcolor=blue,
            urlcolor=cyan,
            citecolor=blue]{hyperref}

\usepackage{trimclip}
\makeatletter
\DeclareRobustCommand{\circbullet}{\mathbin{\vphantom{\circ}\text{\circbullet@}}}
\newcommand{\circbullet@}{%
  \check@mathfonts
  \m@th\ooalign{%
    \clipbox{0 0 0 {\dimexpr\height-\fontdimen22\textfont2}}{$\bullet$}\cr
    $\circ$\cr
  }%
}
\DeclareRobustCommand{\bulletcirc}{\mathbin{\text{\bulletcirc@}}}
\newcommand{\bulletcirc@}{%
  \check@mathfonts
  \m@th\ooalign{%
    \raisebox{\fontdimen22\textfont2}{\clipbox{0 {\fontdimen22\textfont2} 0 0}{$\bullet$}}\cr
    $\circ$\cr
  }%
}
\makeatother

\usepackage[backend=bibtex,style=alphabetic,sorting=ynt,
isbn=false,url=false,eprint=false]{biblatex}
 
\usepackage[ruled,vlined,linesnumbered]{algorithm2e}
\SetKwRepeat{Do}{do}{while}
\SetAlCapHSkip{2.5em} 
\SetAlgoCaptionLayout{centerline}

\makeatletter
\newcommand{\mask}[2]{{\mathpalette\mask@{{#1}{#2}}}}
\newcommand{\mask@}[2]{\mask@@{#1}#2}
\newcommand{\mask@@}[3]{%
  \settowidth{\dimen@}{$\m@th#1#2$}%
  \makebox[\dimen@]{$\m@th#1#3$}%
}
\makeatother

\usepackage{scalerel}
\newcommand\smallsquare{\scaleobj{0.7}{\Box}}

\newcommand{\R}{\mathbb{R}}

\newcommand{\eps}{\varepsilon}

\newcommand{\sdf}{\varphi_{\pm}}

\newcommand{\spd}{{\mathscr{V}}}

\newcommand{\nuext}{{\mathring{\bm{\nu}}}}

\newcommand{\sdist}{{\varphi_{\pm}}}

\newcommand{\sgn}{\textrm{sgn}}
\newcommand\mapsfrom{\mathrel{\reflectbox{\ensuremath{\mapsto}}}}
\newcommand*\mean[1]{\bar{#1}}
\newcommand{\datacov}{\mathcal{C}_{\bm{u}^\star}}
\newcommand{\priorcov}{\mathcal{C}_{\mean{\bm{x}}}}
\newcommand{\postcovmap}{\mathcal{C}_{\bm{x}^\circ}}
\newcommand{\postcovxk}{\mathcal{C}_{\bm{x}_k}}

\newcommand{\Rey}{\mathit{Re}}

\def\dashint{\,\ThisStyle{\ensurestackMath{%
  \stackinset{c}{.2\LMpt}{c}{.5\LMpt}{\SavedStyle-}{\SavedStyle\phantom{\int}}}%
  \setbox0=\hbox{$\SavedStyle\int\,$}\kern-\wd0}\int}

\DeclarePairedDelimiter{\abs}{\lvert}{\rvert}
\DeclarePairedDelimiter{\norm}{\lVert}{\rVert}
\DeclarePairedDelimiter{\normL}{\big\lVert}{\big\rVert}

\DeclarePairedDelimiter{\binner}{\big\langle}{\big\rangle}

\DeclarePairedDelimiterX{\dotp}[2]{\big\langle}{\big\rangle}{#1, #2}
\DeclarePairedDelimiterX{\Dotp}[2]{\Big\langle}{\Big\rangle}{#1, #2}

\font\tenbms=cmbsy10
\font\sevenbms=cmbsy10 at 7pt
\font\fivebms=cmbsy10 at 5pt

\newfam\bmsfam
\textfont\bmsfam=\tenbms
\scriptfont\bmsfam=\sevenbms
\scriptscriptfont\bmsfam=\fivebms

\definecolor{deepblue}{rgb}{0,0,0.5}
\definecolor{deepred}{rgb}{0.6,0,0}
\definecolor{deepgreen}{rgb}{0,0.5,0}
\definecolor{dkgreen}{rgb}{0,0.6,0}
\definecolor{gray}{rgb}{0.5,0.5,0.5}
\definecolor{mauve}{rgb}{0.58,0,0.82}

\newcommand{\REV}[1]{\textcolor{black}{#1}}
\newcommand{\RREV}[1]{\textcolor{black}{#1}}
\newcommand{\RRREV}[1]{\textcolor{black}{#1}}

\newcommand{\REVV}[1]{\textcolor{black}{#1}}

\title{Bayesian inverse Navier--Stokes problems:\\ joint flow field reconstruction and parameter learning}
\author{Alexandros Kontogiannis\textsuperscript{1}$^\dagger$, Scott V. Elgersma\textsuperscript{2},\\ Andrew J. Sederman\textsuperscript{2}, Matthew P. Juniper\textsuperscript{1} \\[.25em] \small \textsuperscript{1}Department of Engineering, University of Cambridge, Trumpington Street, Cambridge CB2 1PZ, UK \\ \small \textsuperscript{2}Department of Chemical Engineering \& Biotechnology, University of Cambridge, \\\small Philippa Fawcett Drive, Cambridge CB3 0AS, UK \\[.25em] \small $^\dagger$Correspondence: \color{blue}{\texttt{ak2239@cam.ac.uk}}}
\date{\today}

\begin{document}

\maketitle

\begin{abstract}
We formulate and solve a Bayesian inverse Navier--Stokes (N--S) problem that assimilates velocimetry data in order to jointly reconstruct a 3D flow field and learn the unknown N--S parameters, including the boundary position. By hardwiring a generalised N--S problem, and regularising its unknown parameters using Gaussian prior distributions, we learn the most likely parameters in a collapsed search space. The most likely flow field reconstruction is then the N--S solution that corresponds to the learned parameters. We develop the method in the variational setting and use a stabilised Nitsche weak form of the N--S problem that permits the control of all N--S parameters. To regularise the inferred geometry, we use a viscous signed distance field (vSDF) as an auxiliary variable, which is given as the solution of a viscous Eikonal boundary value problem. We devise an algorithm that solves this inverse problem, and numerically implement it using an adjoint-consistent stabilised cut-cell finite element method. We then use this method to reconstruct magnetic resonance velocimetry (flow-MRI) data of a 3D steady laminar flow through a physical model of an aortic arch for two different Reynolds numbers and signal-to-noise ratio (SNR) levels (low/high). We find that the method can accurately i) reconstruct the low SNR data by filtering out the noise/artefacts and recovering flow features that are obscured by noise, and ii) reproduce the high SNR data without overfitting. Although the framework that we develop applies to 3D steady laminar flows in complex geometries, it readily extends to time-dependent laminar and Reynolds-averaged turbulent flows, as well as non-Newtonian (e.g. viscoelastic) fluids.
\end{abstract}

\tableofcontents

\section{Introduction}
For inverse problems in flow field imaging (velocimetry), \textit{a priori} knowledge takes the form of a {Navier--Stokes} \mbox{(N--S)} problem. The problem of reconstructing and segmenting a flow image can then be expressed as a generalised inverse Navier--Stokes problem whose flow domain, boundary conditions, and model parameters have to be inferred in order for the modelled velocity (i.e. the reconstruction) to approximate the measured velocity in an appropriate metric space. This approach not only produces a reconstruction that can be made to be an accurate fluid flow inside or around the object, but also provides additional physical knowledge (e.g. pressure, wall shear stress, and effective viscosity field/tensor), which is otherwise difficult or impossible to measure. Inverse N--S problems have been intensively studied during the last decade, mainly enabled by the increase of available computing power. Recent applications in fluid dynamics range from the forcing inference problem \cite{Foures_Dovetta_Sipp_Schmid_2014,Hoang2014}, to the reconstruction of scalar image velocimetry (SIV) \cite{Gillissen2018,Sharma2019} and particle image velocimetry (PIV) \cite{Gillissen2019} data, and the identification of optimal sensor arrangements \cite{Mons2017,Verma2019}. Regularisation methods that can be used to reduce the search space of model parameters are reviewed in \cite{Stuart2010} from a Bayesian perspective, and in \cite{Benning2018} from a variational perspective. \REVV{The well-posedness of Bayesian inverse N--S problems has been addressed in \cite{Cotter2009}}.

\subsection{Flow field reconstruction}

In \cite{Koltukluoglu2018} the authors treat the reduced inverse N--S problem of finding the Dirichlet boundary condition for the inlet velocity that matches the modelled velocity field to flow-MRI data for a steady 3D flow in a glass replica of the human aorta. They measure the data-model discrepancy using the \mbox{$L^2$-norm} and introduce additional variational regularisation terms for the Dirichlet boundary condition. The same formulation is extended to periodic flows in \cite{Koltukluoglu2019,Koltukluoglu2019b}, using the harmonic balance method for the temporal discretisation of the N--S problem. In \cite{Funke2019} the authors address the problem of inferring both the inlet velocity (Dirichlet) boundary condition and the initial condition, for unsteady blood flows and 4D flow-MRI data in a cerebral aneurysm. Note that the above studies consider rigid boundaries and require \textit{a priori} an accurate, and time-averaged, geometric representation of the blood vessel, which is a severe constraint on the inverse problem of finding the velocity field.

To find the shape of the flow domain, e.g. the blood vessel boundaries, computed tomography (CT) or magnetic resonance angiography (MRA) is often used. The acquired image is then reconstructed, segmented, and smoothed. This process not only requires substantial effort and the design of an additional experiment (e.g. CT, MRA), but it also introduces geometric uncertainties \cite{Morris2016,Sankaran2016}, which, in turn, affect the predictive confidence of arterial wall shear stress distributions and their mappings \cite{Katritsis2007,Sotelo2016}. For example, in \cite{Funke2019} the authors report discrepancies between the modelled and the measured velocity fields near the flow boundaries, and they suspect they are caused by geometric errors that were introduced during the boundary segmentation process. In general, the assumption of rigid boundaries either implies that a time-averaged geometry has to be used, or that an additional experiment (e.g. CT, MRA) has to be conducted to register the moving boundaries to the flow measurements. A more consistent approach to this problem is to treat the blood vessel geometry as an unknown when solving the generalised inverse N--S problem \cite{Kontogiannis2021}. In this way, the method simultaneously reconstructs and segments the velocity fields and can better adapt to the velocimetry experiment by correcting the geometric errors and improving the reconstruction. In other words, the velocity field is used to find the boundary, and the boundary is used to find the velocity field. This is a neater optimisation problem since it removes the severe constraint of fixing the boundary before assimilating the velocity field.

In this study we build upon the work in \cite{Kontogiannis2021}. \RRREV{In particular, we i) extend the methodology from 2D to 3D flows, ii) define the data-to-model projection operator for imaging problems, iii) incorporate a stabilised Nitsche cut-cell finite element method for the N--S problem that remains robust for high $\Rey$ numbers and leads to an adjoint-consistent discrete formulation (i.e. the discretised continuous adjoint operator is equivalent to the discrete adjoint operator), iv) revise and improve the implicit representation of the geometry by incorporating the viscous Eikonal equation as an additional constraint to the problem, and v) describe the numerical problem in detail by defining the discrete function spaces and the stabilised discrete weak forms that make up the inverse problem. We then devise an improved algorithm that solves this Bayesian inverse N--S problem, and demonstrate it on flow-MRI data of a 3D steady laminar flow through a physical model of an aortic arch for low and high $\Rey$ numbers.}

\subsubsection{Deep learning reconstruction algorithms}
Artificial intelligence (AI) algorithms, such as deep neural networks (NNs), are versatile, relatively simpler to implement, and have revolutionised the field of computer vision. There are, however, fundamental problems that still need to be addressed for their application to physics-based problems such as flow field reconstruction (e.g. AI-generated hallucinations, and the problem of existence of stable computational algorithms) \cite{9420272,doi:10.1073/pnas.2107151119}. Physics-informed NNs (PINNs) cannot yet solve boundary value problems, (e.g the N--S problem) more efficiently than finite element methods \cite{grossmann2023physicsinformed}. Further, PINNs contain orders of magnitude more degrees of freedom than the method in this paper, with a corresponding increase in the required amounts of data and training time. The method in this paper is, in contrast, formulated from a N--S boundary value problem in a variational framework. The physics is \emph{hardwired} into the model, meaning that the search space is restricted to solutions that satisfy the \mbox{N--S} problem. This is a key-difference between PINNs and the approach in this study: while PINNs solve a minimisation problem\footnote{\REV{It is possible to hardwire constraints with PINNs \cite{doi:10.1137/21M1397908}, but then the following question arises: do PINNs make for better PDE-solvers? The answer is no \cite{grossmann2023physicsinformed}. In fact, for inverse Navier--Stokes problems with hard constraints, the success of the method heavily relies on the robustness of the forward/adjoint solvers.}}, we solve a saddle point problem. In other words, PINNs do not hardwire the N--S problem; they simply penalise its residuals (i.e. they treat the N--S problem as a soft constraint). In our experience, solving the saddle point problem, even though more computationally expensive, leads to superior flow field reconstructions and more accurate inferred parameters. Unlike NNs \cite{Ferdian2020,Rutkowski2021} and PINNs \cite{SAITTA2024108057,Zhu_Jiang_Lefauve_Kerswell_Linden_2024}, our model is relatively small, physically interpretable, amenable to mathematical analysis, extrapolatable, and will reconstruct flows that it has not seen before, thus enabling digital twin applications in engineering design and patient-specific cardiovascular modelling.

\subsection{Outline}
\begin{itemize}
\item In section 2 we present the general framework for the Bayesian inversion of nonlinear models.
\item In section 3 we formulate the Bayesian inversion of the Navier--Stokes problem, in a continuous, variational setting, and devise an algorithm that solves it.
\item In section 4 we formulate the discrete Bayesian inverse N--S problem, and discuss the details of its numerical implementation.
\item In section 5 we use the algorithm to reconstruct flow-MRI data of a 3D steady laminar flow through a physical model of an aortic arch.
\end{itemize}

\subsection{Notation}
In what follows, $L^2(\Omega)$ denotes the space of square-integrable functions in $\Omega$, with inner product $\binner{\cdot,\cdot}$ and norm $\norm{\cdot}_{L^2(\Omega)}$, and $H^k(\Omega)$ denotes the space of square-integrable functions with $k$ square-integrable derivatives. For a given covariance operator, $\mathcal{C}$, we also define the covariance-weighted $L^2(\Omega)$ spaces, endowed with the inner product ${\binner{\cdot,\cdot}_{\mathcal{C}(\Omega)} := \binner{\mathcal{C}^{-1/2}\cdot,\mathcal{C}^{-1/2}\cdot}}_{\Omega}$, which generates the norm $\norm{\cdot}_{\mathcal{C}(\Omega)}$. The Euclidean norm in the space of real numbers $\R^n$ is denoted by $\abs{\cdot}_{\R^n}$, and the measure (volume) of the domain $\Omega$ by $\abs{\Omega}$. The first variation of a functional $\mathscr{J}:L^2(\Omega)\to\mathbb{R}$ is defined by
\begin{equation}
\delta_{z}\mathscr{J}\equiv\frac{d}{d\tau}\mathscr{J}({z}+\tau{z}')\Big\vert_{\tau=0}=\binner{D_z\mathscr{J},z'}_\Omega\quad,
\label{eq:first_variation}
\end{equation}
where $\tau \in \mathbb{R}$, $z\in L^2(\Omega)$, $z'\in L^2(\Omega)$ is an allowed perturbation of $z$, and $D_z\mathscr{J}$ is the generalised gradient. We use the superscript $(\cdot)^*$ to denote the adjoint of an operator, $(\cdot)^\star$ to denote a measurement, and $(\cdot)^\circ$ to denote a reconstruction. For instance, $\bm{u}^\star$ denotes the measured velocity obtained from a velocimetry experiment, $\bm{u}^\circ$ denotes the corresponding reconstructed velocity field, and $\bm{u}$ denotes a (modelled) velocity field obtained from the N--S problem.

\section{Bayesian inversion of nonlinear models}
\label{sec:bayesian_inv}

Given experimental, possibly noisy and sparse, vector field data of a physical quantity, $\bm{u}^\star$, we assume that we already know a physical model, $\bm{u}=\bm{u}(\bm{x})$, that can fit the data for suitable, \emph{physically-interpretable} parameters, $\bm{x}$. In other words, there exist parameters $\bm{x}^\circ$ such that 
\begin{gather}
\bm{u}^\star \simeq \mathcal{Z}\bm{x}^\circ = \big(\mathcal{S}\mathcal{Q}\big)\bm{x}^\circ = \mathcal{S}\bm{u}^\circ \quad,
\label{eq:model_approx_data}
\end{gather}
where \RREV{`$\simeq$' denotes an approximation in some metric space to be defined later}, $\mathcal{Z}\coloneqq\mathcal{S}\mathcal{Q}$ is the operator that maps model parameters to model solutions projected into the data space, $\mathcal{S}$ is the model space to data space projection operator, which is linear and non-invertible\footnote{Except if the two spaces are identical, in which case $\mathcal{S}\equiv\mathrm{I}$, where $\mathrm{I}$ is the identity operator.}, and $\mathcal{Q}$ is the operator that encapsulates the physical model, which is nonlinear. Under this assumption, the main goal is to find $\bm{x}^\circ$, which is of interest because it often pertains to physical quantities that cannot be directly measured, and which are instead inferred from the data. At the same time, we obtain the reconstructed (model-filtered) velocity field, $\bm{u}^\circ\equiv\mathcal{Q}\bm{x}^\circ$, which fits \RREV{(as opposed to overfits)} \REVV{the noisy, sparse, and under-resolved data, $\bm{u}^\star$}. Starting from the assumptions that i) the \emph{model explains the data}, and ii) the \emph{data noise is Gaussian}, we write
\begin{gather}
\bm{u}^\star - \mathcal{Z}\bm{x} = \bm{\eps} \sim \mathcal{N}(\bm{0},\datacov)\quad,
\end{gather}
where $\bm{\eps}\sim\mathcal{N}(\bm{0},\datacov)$ is Gaussian noise with zero mean and with covariance operator $\datacov$\footnote{In addition to data noise, it is possible to account for model errors by incorporating an additional term $\bm{\eps}_M$, which \RREV{can be} approximated by another Gaussian distribution.}. We also assume that we have prior knowledge of the probability distribution of the model parameters, $\bm{x}$. We call this the prior parameter distribution, $\mathcal{N}(\mean{\bm{x}},\priorcov)$, which we assume to be Gaussian with prior mean $\mean{\bm{x}}$, and prior covariance operator $\priorcov$. We then use Bayes' theorem, which states that the posterior probability density function (p.d.f.) of $\bm{x}$, given the data $\bm{u}^\star$, $\pi\big(\bm{x}\big|\bm{u}^\star\big)$, is proportional to the data likelihood, $\pi\big(\bm{u}^\star\big|\bm{x}\big)$, times the prior p.d.f. of $\bm{x}$, $\pi\big(\bm{x}\big)$, i.e.
\begin{align}
\pi\big(\bm{x}\big|\bm{u}^\star\big) &\propto \pi\big(\bm{u}^\star\big|\bm{x}\big)~\pi(\bm{x}) \nonumber\\
&= \exp\Big(-\frac{1}{2}\norm{\bm{u}^\star-\mathcal{Z}\bm{x}}^2_{\datacov} 
- \frac{1}{2}\norm{\bm{x}-\mean{\bm{x}}}^2_{\priorcov}\Big)\label{eq:posterior_pdf_0}
\end{align}
where $\pi(\cdot)$ is the Gaussian p.d.f., and $\norm{\cdot,\cdot}^2_{\mathcal{C}} \coloneqq \binner{\cdot,\mathcal{C}^{-1}\cdot}$ is the covariance-weighted $L^2$-norm. The most likely parameters $\bm{x}^\circ$, in the sense that they maximise the posterior p.d.f. (maximum \emph{a posteriori} probability, or MAP estimator), are given implicitly as the solution of the nonlinear optimisation problem
\begin{gather}
\RREV{\bm{x}^\circ \equiv \underset{\bm{x}}{\mathrm{argmin}}\mathscr{J}(\bm{x}) \quad,\quad \text{where}\quad \mathscr{J}(\bm{x}) \coloneqq \frac{1}{2}\norm{\bm{u}^\star-\mathcal{Z}\bm{x}}^2_{\datacov} 
+ \frac{1}{2}\norm{\bm{x}-\mean{\bm{x}}}^2_{\priorcov}\quad.}
\label{eq:map_opt_problem}
\end{gather}

\subsection{Nonlinearity and optimisation}
\label{sec:nonlinearity_bayes}
In order to solve problem \eqref{eq:map_opt_problem} we first linearise $\bm{u}$ around $\bm{x}_k$, i.e.
\begin{gather}
\bm{u} = \mathcal{Q}\bm{x} \simeq \mathcal{Q}\bm{x}_k + \mathcal{A}_k\big(\bm{x}-\bm{x}_k)\quad, 
\label{eq:def_of_a}
\end{gather}
where $\mathcal{A}_k \equiv \big(D^{\bm{u}}_{\bm{x}}\big)_k$ is the Jacobian of $\bm{u}$ with respect to $\bm{x}$. In the same spirit, we obtain
\begin{gather}
\mathcal{Z}\bm{x} \simeq \mathcal{Z}\bm{x}_k + \mathcal{G}_k\big(\bm{x}-\bm{x}_k)\quad,\quad \mathcal{G}_k \coloneqq \mathcal{S}\mathcal{A}_k\quad.
\label{eq:z_taylor_exp}
\end{gather}
\REVV{To find the unknown $\bm{x}$ that satisfies the optimality conditions of \eqref{eq:map_opt_problem}, we can use an iteration based on the following update rule \cite[Chapter~6.22.6]{Tarantola2005}}
\begin{gather}
\bm{x}_{k+1} \mapsfrom \bm{x}_k - \tau_k~\postcovxk~\big(D_{\bm{x}}\mathscr{J}\big)_k\quad,
\label{eq:map_update_rule}
\end{gather}
where $\mathbb{R} \ni \tau_k > 0$ is the step size at iteration $k$, which is determined by a line search algorithm, $\postcovxk$ is the posterior (parameter) covariance operator at iteration $k$, which is given by
\begin{gather}
\postcovxk \coloneqq \big(\mathcal{G}_k^*~\datacov^{-1}~\mathcal{G}_k+\priorcov^{-1}\big)^{-1}\quad,
\label{eq:map_cov}
\end{gather}
where $\mathcal{G}^{*}_k$ is the adjoint of $\mathcal{G}_k$, and $\big(D_{\bm{x}}\mathscr{J}\big)_k$ is the gradient of the objective, $\mathscr{J}$, with respect to the parameters, $\bm{x}$, which is given by
\begin{gather}
\big(D_{\bm{x}}\mathscr{J}\big)_k \coloneqq \underbrace{-\mathcal{G}_k^*~\datacov^{-1}\big(\bm{u}^\star-\mathcal{Z}\bm{x}_k\big)}_{\text{model term}} + \underbrace{\priorcov^{-1}\big(\bm{x}_k-\bar{\bm{x}}\big)}_{\text{prior term}}\quad.
\label{eq:map_update_gradient}
\end{gather}

\subsection{Posterior distribution and the Gaussian assumption}
\label{sec:laplace_approx}
\begin{figure}[!h]
\centering
\includegraphics[width=\textwidth]{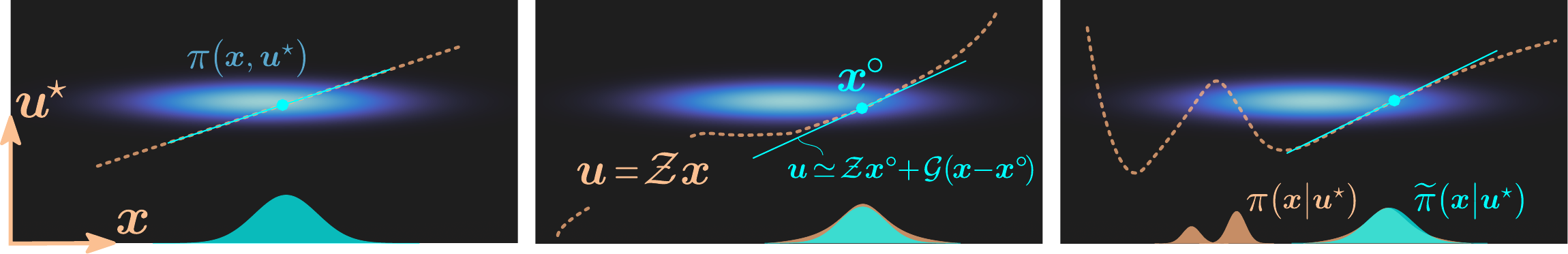}
\caption{When $\mathcal{Z}$ is linear (left), $\pi(\bm{x}|\bm{u}^\star)$ is Gaussian. When $\mathcal{Z}$ is weakly nonlinear (middle), $\pi(\bm{x}|\bm{u}^\star)$ can be approximated reasonably well by a Gaussian p.d.f., $\widetilde{\pi}(\bm{x}|\bm{u}^\star)$,  around $\bm{x}^\circ$ (Laplace approximation). When $\mathcal{Z}$ is strongly nonlinear (right), $\pi(\bm{x}|\bm{u}^\star)$ can even be multimodal, in which case there are multiple critical points which will yield different approximations $\widetilde{\pi}(\bm{x}|\bm{u}^\star)$ (even locally, the approximation may be inaccurate).}
\label{fig:ls_nonlinearity}
\end{figure}

For $\bm{x}$ in a neighbourhood of the MAP point, $\bm{x}^\circ$, we can furthermore write
\begin{equation}
\pi\big(\bm{x}\big|\bm{u}^\star\big)\simeq  \widetilde{\pi}\big(\bm{x}\big|\bm{u}^\star\big) \coloneqq \exp\Big(-\frac{1}{2}\norm{\bm{x}-\bm{x}^\circ}^2_{\postcovmap} - \textrm{const.}\Big)\quad,
\label{eq:laplace_approx}
\end{equation}
which is known as the \emph{Laplace approximation} \cite[Chapter~27]{MacKay2003}, and is exact when $\mathcal{Z}$ is a linear operator. Figure \ref{fig:ls_nonlinearity} illustrates the conceptual difference between linear, weakly nonlinear, and strongly nonlinear operators $\mathcal{Z}$. In this study, $\mathcal{Z}$ encodes a Navier--Stokes problem, and the `strength' of the nonlinearity depends on the flow conditions (e.g. on the geometry and the boundary conditions) and the Reynolds number, $\Rey$. Note that, for $\Rey \ll 1$, the N--S problem can be approximated well enough by the Stokes problem, which is linear. At the other end of the spectrum, for sufficiently large $\Rey$, the flow becomes turbulent, which is a strongly nonlinear phenomenon. It should be emphasised that what matters is the behaviour of $\mathcal{Z}$ around the MAP point, $\bm{x}^\circ$, and within the $\sim3\sigma$ support of the joint density, $\pi(\bm{x},\bm{u}^\star)$. If $\mathcal{Z}$ is to be linearised around $\bm{x}^\circ$, the greater the spread of $\pi(\bm{x},\bm{u}^\star)$, the less accurate the approximation will be. Also, the spread of $\pi(\bm{x},\bm{u}^\star)$ depends on i) the noise level in the data, and ii) the assumed prior distribution of the N--S unknowns, $\bm{x}$. This implies that, when $\mathcal{Z}$ is strongly nonlinear, we may require higher-quality data and/or more informative priors in order to regularise the inverse problem. Alternatively, we can simplify or reformulate the model such that $\mathcal{Z}$ exhibits smoother behaviour. In turbulent flows, for example, the large scales will depend on the geometry and the boundary conditions of the problem, while the small scales of turbulence exhibit a universal behaviour that is not directly linked to these parameters. Hence, inferring the evolution of the averaged velocity and pressure fields using the \emph{Reynolds-averaged} Navier--Stokes (RANS) equations, instead of the instantaneous fields that are governed by the N--S equations, can lead to a well-posed inverse problem. In this paper, however, we only consider laminar flow with a Newtonian viscosity model.   

\subsection{Posterior covariance approximation}
\label{sec:uncertainty_estimation}
Instead of using formula \eqref{eq:map_cov} to compute the exact posterior parameter covariance operator, $\postcovxk$, which is computationally expensive, we can approximate it using a quasi-Newton method (e.g. BFGS) \cite{Fletcher2000}. Quasi-Newton methods use the states, $\bm{x}_k$, and the gradients, $\big(D_{\bm{x}}\mathscr{J}\big)_k$, to reconstruct the Hessian operator or its inverse. In the case of the quadratic problem \eqref{eq:map_opt_problem}, the inverse Hessian operator, $H_k$, is identical to $\postcovxk$.
\RREV{The reconstructed inverse Hessian, $\widetilde{H}_k$, which is equivalent to the reconstructed posterior (parameter) covariance, $\widetilde{\mathcal{C}}_k$, is then given by the recursive formula \cite{Yan1996}\cite[Chapter~6.22.8]{Tarantola2005}}
\REVV{\begin{align}
\widetilde{\mathcal{C}}_k~\cdot = \priorcov~\cdot &+ \sum_{j=1}^{k-1}\big(1 + \omega_j\rho_j \big)\rho_j~\binner{\bm{s}_j,\cdot}~\bm{s}_j\nonumber\\
&+\sum_{j=1}^{k-1}-\rho_j~\binner{\cdot,\bm{\alpha}_j}~\bm{s}_j - \rho_j~\binner{\bm{s}_j,\cdot}~\bm{\alpha}_j\quad,
\label{eq:bfgs_recursive_formula}
\end{align}}
where 
\begin{gather}
\bm{s}_k \mapsfrom \bm{x}_k-\bm{x}_{k-1} \quad\text{and}\quad \bm{y}_k \mapsfrom \big(D_{\bm{x}}\mathscr{J}\big)_k-\big(D_{\bm{x}}\mathscr{J}\big)_{k-1}\quad,
\end{gather}
$\cdot$ is a placeholder for the argument, $\omega_j \equiv \binner{\bm{y}_j,\widetilde{\mathcal{C}}_j\bm{y}_j}$, $\rho_j \equiv 1/\binner{\bm{y}_j,\bm{s}_j}$, and $\bm{\alpha}_j \equiv \widetilde{\mathcal{C}}_j\bm{y}_j$, are quantities that are computed only once and stored in memory. Note that we here choose to initialise the posterior covariance using the prior covariance, i.e. $\mathcal{C}_0 \mapsfrom \priorcov$, but the best initialisation strategy depends on the specifics of the problem. 

\subsubsection{Uncertainty estimation}
\label{sec:uncert_estim}
To estimate the predicted uncertainty around the MAP solution, we can use Arnoldi iteration to compute the eigendecomposition of either the exact posterior covariance, which is given by \eqref{eq:map_cov}, or the approximated posterior covariance, which is given by \eqref{eq:bfgs_recursive_formula}. A third option is to reconstruct $\postcovmap$ using the eigendecompositions of $\mathcal{G}_{\bm{x}^\circ}$, $\mathcal{G}^*_{\bm{x}^\circ}$, $\datacov^{-1}$, $\priorcov^{-1}$, which may be simpler to compute. Once the eigenvalue/eigenvector pairs, $(\lambda,\bm{v})_\ell$,  of the covariance operator are obtained, we find
\begin{equation}
\REVV{\postcovmap~\cdot = \sum^\infty_{\ell} \lambda_\ell~ \bm{v}_\ell~\binner{\bm{v}_\ell,\cdot}\quad.}
\end{equation}
The pointwise variance, $\textrm{var}(\bm{x})$, which is often used to quantify uncertainty, is then given by
\begin{equation}
\mathrm{var}(\bm{x}) = \sum^\infty_\ell \lambda_\ell~\bm{v}^2_\ell\quad,
\end{equation}
and the total variance is given by
\begin{equation}
\norm{\bm{x}-\bm{x}^\circ}^2_{\postcovmap} = \sum^\infty_\ell \lambda_\ell\quad.
\end{equation}
Samples, $\bm{x}^s$, can be drawn using the \mbox{Karhunen--Lo\`eve} (K--L) expansion
\begin{equation}
\bm{x}^s = \bm{x}^\circ + \sum^\infty_\ell \eta_\ell~\sqrt{\lambda_\ell}~\bm{v}_\ell\quad, \quad \text{with}\quad \mathbb{R}\ni\eta_\ell \sim \mathcal{N}(0,1)\quad,
\label{eq:kl_expansion}
\end{equation}
in order to further inspect the posterior distribution around the MAP point. Likewise, the posterior covariance of the model solution at iteration $k$, which is given by
\begin{equation}
\mathcal{C}_{\bm{u}_k} \coloneqq \mathcal{A}_{\bm{x}_k}{\mathcal{C}}_{\bm{x}_{k}}\mathcal{A}_{\bm{x}_k}^*\quad,\label{eq:state_covariance}
\end{equation}
can be reconstructed from the eigendecompositions of $\postcovxk$, $\mathcal{A}_k$, and $\mathcal{A}^*_k$.

\begin{figure}
\centering
\includegraphics[width=\textwidth]{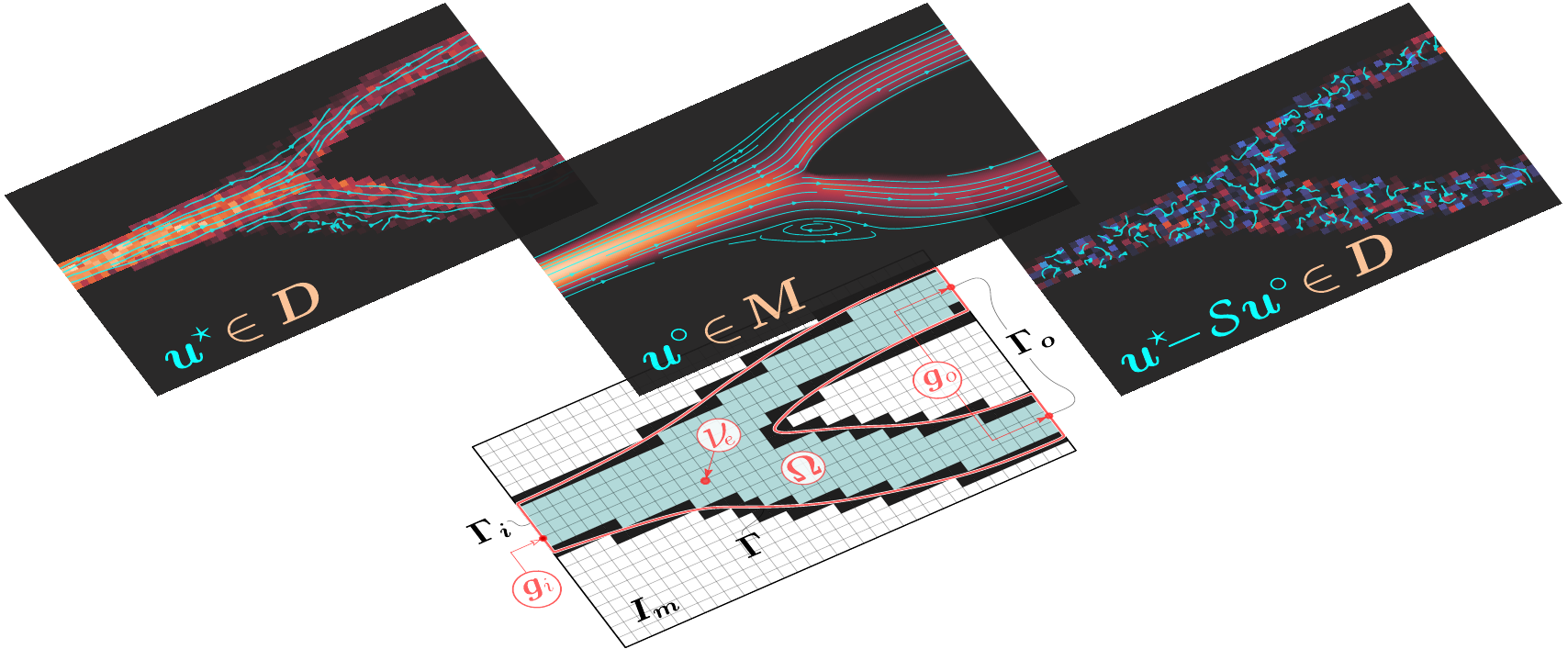}
\caption{The Bayesian inverse N--S problem assimilates velocimetry data, $\bm{u}^\star$, in order to jointly reconstruct the flow field, $\bm{u}$, and learn the unknown N--S parameters, $\bm{x}$. The MAP estimator of the unknown N--S parameters, $\bm{x}^\circ$, is found by solving problem \eqref{eq:inv_prob_aug}, and the reconstructed velocity field is given by $\bm{u}^\circ=\mathcal{Q}\bm{x}^\circ$.}
\label{fig:flow_reconstruction_concept}
\end{figure}

\section{Bayesian inversion of the Navier--Stokes problem}
\label{sec:bayesian_inv_ns_formulation}
In section \ref{sec:bayesian_inv} we discussed the general Bayesian inversion problem and the MAP estimator, given by \eqref{eq:map_opt_problem}, which can be iteratively solved by using the update rule \eqref{eq:map_update_rule} along with the objective gradients \eqref{eq:map_update_gradient} and the posterior covariance \eqref{eq:map_cov} (or its BFGS-approximation) when the operators $\datacov$, $\priorcov$, $\mathcal{S}$, $\mathcal{G}$, and $\mathcal{Q}$ are precisely defined. In order to obtain these operators for the Bayesian inversion of the generalised Navier--Stokes problem, we need to derive the first order optimality conditions of the following \emph{saddle point} problem
\begin{equation}
\text{find} \quad \bm{x}^\circ \equiv \mathrm{arg}\ \underset{\bm{x}}{\mathrm{min}}\ \underset{\bm{v},q,r}{\mathrm{max}}~\mathscr{J}(\bm{u},p,\bm{v},q,r;\bm{x})\quad,
\label{eq:inv_prob_aug}
\end{equation}
where
\begin{gather}
\RREV{{\mathscr{J}}(\bm{u},p,\bm{v},q,r;\bm{x}) \coloneqq {\mathscr{U}}(\bm{u}) + \mathscr{R}_{\bm{x}}(\bm{x}) + {\mathscr{M}}(\bm{u},p,\bm{v},q;\bm{x}) + \mathscr{D}(r,\sdf) \quad.
\label{eq:obj_func}}
\end{gather}
The augmented objective functional, $\mathscr{J}$, incorporates the data-model discrepancy, $\mathscr{U}$, the priors of the N--S parameters, $\mathscr{R}_{\bm{x}}$, and two nonlinear equality constraints: i) the generalised Navier--Stokes boundary value problem, $\mathscr{M}$, for $(\bm{u},p)$ (i.e. velocity and pressure), and ii) the viscous Eikonal boundary value problem, $\mathscr{D}$, for the viscous signed distance field (vSDF), $\sdist$. The vSDF, $\sdist$, is an auxiliary variable that we use to implicitly define and control the geometry. The Lagrange multipliers, $(\bm{v},q,r)$, also known as adjoint variables, are used to enforce the two model constraints: i) $(\bm{v},q)$ enforce the N--S b.v.p. for $(\bm{u},p)$, and ii) $r$ enforces the vSDF b.v.p. for $\sdist$. \REVV{Note that, problem \eqref{eq:inv_prob_aug}-\eqref{eq:obj_func} is the Lagrangian formulation of the MAP estimation problem \eqref{eq:map_opt_problem}, where the  nonlinear constraint \mbox{$\bm{u} = \mathcal{Q}\bm{x}$}, which is now encoded in $\mathscr{M}$, is enforced using Lagrange multipliers. The additional constraint on the vSDF, which is encoded in $\mathscr{D}$, is because we use $\sdist$ as an auxiliary variable to define the geometry, $\Omega$, which belongs to the N--S unknowns, $\bm{x}$.}

In this section, we first define $\datacov$ and $\mathcal{S}$, then we use variational calculus to construct $\mathcal{G}$ and $\mathcal{Q}$, and lastly we define $\priorcov$. We further devise an algorithm that solves the Bayesian inverse N--S problem, which is presented at the end of this section.

\subsection{Data-model discrepancy: operators $\mathcal{S}$, \texorpdfstring{$\datacov$}{}}
\begin{figure}
\centering
\includegraphics[width=\textwidth]{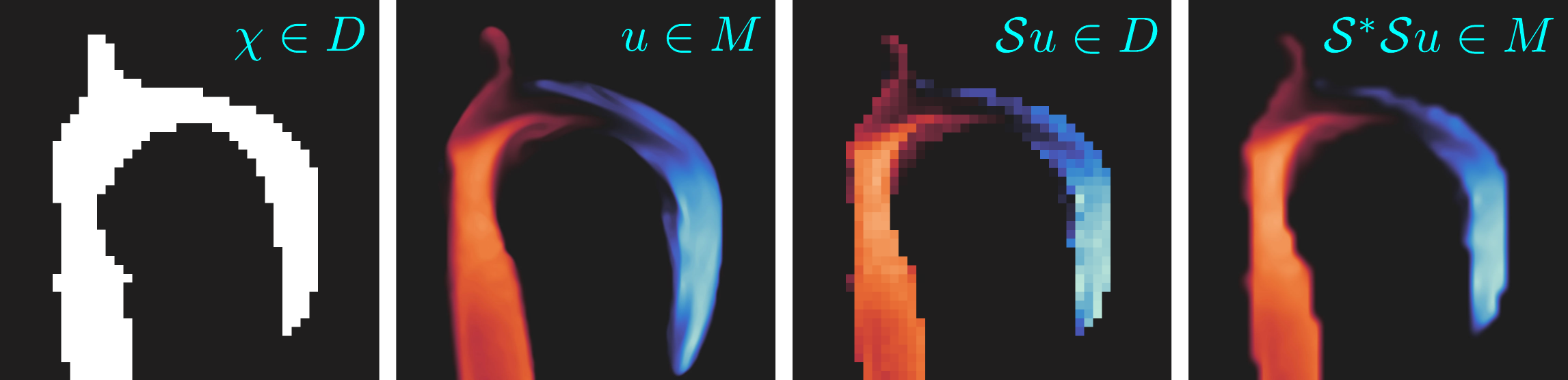}
\caption{The projection operator $\mathcal{S}: \bm{M}\to\bm{D}$ is used to project the modelled velocity field, $\bm{u} \in \bm{M}$, to $\mathcal{S}\bm{u}\in\bm{D}$, so that it can be compared with the velocity data, $\bm{u}^\star$. A mask, $\chi$, is commonly used after pre-processing the data to remove irregularities such as outliers (e.g. defective pixels), or regions with no signal. The adjoint operator $\mathcal{S}^*$ is used to project from $\bm{D}$ to $\bm{M}$ (the inverse of $\mathcal{S}$ does not exist).}
\label{fig:s_proj}
\end{figure}
We define the data space $\bm{D} \equiv \bm{L}^2(I_d)$, and the model space $\bm{M}\equiv \bm{L}^2(I_m)$, where $I_d$, $I_m$ are bounded subsets of $\R^3$. The size of $I_d$ depends on the field of view (FOV) of the experiment, and the size of $I_m$ depends on the field of view of the model. For problems in velocity imaging, such as flow-MRI, the data domain, $I_d$, is a uniform tessellation consisting of $N$ (disjoint) voxels (or pixels in 2D), $\{V_i\}_{i=1}^N$, such that 
\begin{equation}
I_d = \bigcup_{i=1}^N V_i\quad,\quad\text{and}\quad \textrm{int}(V_i)\cap \textrm{int}(V_j)=\varnothing\quad \text{for}~ i\neq j\quad,
\end{equation}
where $\textrm{int}(\cdot)$ denotes the interior of a set. The data space, $\bm{D}$, is the function space on $I_d$ that is comprised of all piecewise constant functions
\begin{equation}
\bm{u}_d(x) = \sum_{i=1}^N \bm{\varphi}_i(x)~{\bm{u}^i_d} \quad,
\end{equation}
where $\bm{\varphi}_i(x) \equiv (1,1,1)$ for $x\in V_i$ and $\bm{\varphi}_i(x)\equiv\bm{0}$ otherwise, and $\bm{u}^i_d\in\R^3$. In general, the model space, $\bm{M}$, is a subspace of $\bm{L}^2(I_m)$, but more can be said about $\bm{M}$ and $I_m$ after discretising the N--S problem, which we will address in section \ref{sec:stab_cutcell_fem}. The data-model discrepancy is measured on the data space, and is given by
\begin{gather}
\mathscr{U}(\bm{u}) \coloneqq \frac{1}{2}~\normL{\bm{u}^\star-\mathcal{S}\bm{u}}^2_{{\datacov}}\quad,
\end{gather}
where $\bm{u}=(u_x,u_y,u_z)$, $\bm{u}^\star = (u^\star_x,u^\star_y,u^\star_z)$, and $\mathcal{S}:\bm{M}\to\bm{D}$ is the model-to-data projection operator such that
\begin{equation}
\label{eq:model_to_data_proj}
\mathcal{S}\bm{u}_m \coloneqq \sum_{i=1}^N\chi_i~\bigg(\frac{1}{\abs{V_i}}\int_{V_i}\bm{u}_m\bigg)~\bm{\varphi}_i(x)\quad,\quad \bm{u}_m \in \bm{M}\quad,
\end{equation}
where $\chi = \sum_{i=1}^N\chi_i\varphi_i(x)$ is a mask. Then, \RREV{by definition}, the adjoint operator $\mathcal{S}^*$ satisfies
\begin{equation}
\dotp{\bm{u}_d}{\mathcal{S}\bm{u}_m}_{\bm{D}} = \dotp{\mathcal{S}^*\bm{u}_d}{\bm{u}_m}_{\bm{M}} \quad,\quad \text{for all}\quad \bm{u}_m \in \bm{M},~ \bm{u}_d\in\bm{D}\quad.
\end{equation}
We further assume that the noise colour in the velocity images is white, and define the covariance operator
\begin{equation}
\datacov = \mathrm{diag}\big(\sigma^2_{u^\star_x}\mathrm{I},~\sigma^2_{u^\star_y}\mathrm{I},~\sigma^2_{u^\star_z}\mathrm{I}\big)\quad,
\end{equation}
where $\sigma_{u^\star_x}^2,\sigma_{u^\star_y}^2,\sigma_{u^\star_z}^2$ are the variances of $u^\star_x,u^\star_y,u^\star_z$, respectively, and \RREV{$\mathrm{I}:L^2(I_d)\to L^2(I_d)$} is the identity operator. \REV{The Gaussian white noise assumption is valid for velocity images obtained from phase-contrast MRI when the signal-to-noise ratio (SNR) is above 3 and the frequency ($\bm{k}$-) space is fully sampled \cite{Gudbjartsson1995}.} If the noise is Gaussian but coloured (i.e. correlated), $\datacov$ has to be revised \cite{Kontogiannis2022b}. Having defined $\mathcal{S}$, $\mathcal{S}^*$, and $\datacov$, the first variation of $\mathscr{U}$ with respect to the velocity field is then given by
\begin{align}
\delta_{\bm{u}}\mathscr{U} \coloneqq ~& \dotp{D_{\bm{u}}\mathscr{U}}{\bm{u}'}_{I_m}\quad,\quad\text{where}\quad D_{\bm{u}}\mathscr{U}\equiv -\mathcal{S}^\ast\datacov^{-1}(\bm{u}^\star-\mathcal{S}\bm{u})\quad.
\end{align}
In what follows we will show that the discrepancy, $D_{\bm{u}}\mathscr{U} \in \bm{M}$, is the forcing term of the adjoint N--S problem, whose solution plays an important role in obtaining the model term of the objective gradient \eqref{eq:map_update_gradient}.

\subsection{Generalised N--S problem: operators $\mathcal{G}$, $\mathcal{Q}$}
\label{sec:gen_ns_problem}
The generalised Navier--Stokes boundary value problem (b.v.p.) in $\Omega \subset I_m \subset \R^3$ is given by
\begin{equation}
  \left\{\begin{alignedat}{2}
    \bm{u}\bm{\cdot}\nabla\bm{u}-\nabla\bm{\cdot}\big(2{\color{deepred}\nu_e}\nabla^s\bm{u}\big)  + \nabla p &= \bm{0} \quad &&\textrm{in}\quad {\color{deepred}\Omega} \\
    \nabla \bm{\cdot} \bm{u} &= 0 \quad &&\textrm{in}\quad {\color{deepred}\Omega}\\
    \bm{u} &= \bm{0} \quad &&\textrm{on}\quad {\color{deepred}\Gamma} \\
    \bm{u} &= T_{\Gamma_i}~{\color{deepred}\bm{g}_i} \quad &&\textrm{on}\quad {\color{deepred}\Gamma_i} \\
   -2{\color{deepred}\nu_e}\nabla^s\bm{u}\bm{\cdot}\bm{\nu}+p\bm{\nu} &= T_{\Gamma_o}~{\color{deepred}\bm{g}_o} \quad &&\textrm{on}\quad {\color{deepred}\Gamma_o}
  \end{alignedat}\right. \quad,\quad \underbrace{{\color{black}\bm{x}} = ({\color{deepred}\bm{g}_i},{\color{deepred}\bm{g}_o},{\color{deepred}\nu_e},{\color{deepred}\Omega})}_{\text{unknown parameters}}\quad,
  \label{eq:navierstokes_bvp}
\end{equation}
where $\bm{u}$ is the velocity, $p\mapsfrom p/\rho$ is the reduced pressure, $\rho$ is the density, $\nu_e$ is the \emph{effective} (kinematic) viscosity, $\nabla^s\bm{u} \equiv (\nabla^s \bm{u})_{ij} \coloneqq \frac{1}{2}(\partial_j u_i + \partial_i u_j)$ is the strain-rate tensor, $\bm{g}_i$ is the Dirichlet boundary condition (b.c.) at the inlet $\Gamma_i$, $\bm{g}_o$ is the natural b.c. at the outlet $\Gamma_o$, and $\bm{\nu}$ is the unit normal vector on the boundary $\partial\Omega = \Gamma\cup\Gamma_i\cup\Gamma_o$, where $\Gamma$ is the part of the boundary where a no-slip b.c. is imposed. The trace operators, $T_{\Gamma_i}$ and $T_{\Gamma_o}$, are defined in appendix \ref{app:ext_trace_ops}. We call this N--S b.v.p. `generalised' because all of its input parameters, namely the shape of the domain $\Omega$, the boundary conditions $\bm{g}_i, \bm{g}_o$, and the viscosity field $\nu_e$, are considered to be unknown. Consequently, problem \eqref{eq:navierstokes_bvp} can explain many different wall-bounded incompressible flows for Newtonian or generalised Newtonian fluids (i.e. `power-law' non-Newtonian fluids).

\subsubsection{Weak form using Nitsche's method}
To construct $\mathcal{Q}$, $\mathcal{G}$, we start from the weak form of the Navier--Stokes problem, $\mathscr{M}$, for velocity test functions \mbox{$\bm{v} \in \bm{H}^1(\Omega)$}, and pressure test functions \mbox{$q\in L^2(\Omega)$}. We weakly enforce Dirichlet (essential) velocity boundary conditions on $\Gamma_e \coloneqq \Gamma \cup \Gamma_i$ using Nitsche's method \cite{Nitsche1971}, which was initially formulated for the Poisson problem and later on extended to the linearised N--S (Oseen) problem \cite{Burman2006,Bazilevs2007,Massing2018}. This is done by augmenting $\mathscr{M}$ with the following terms
\begin{subequations}
\begin{align}
\mathscr{N}_\Delta(\bm{v},q,\bm{u};\bm{g}_i) &\coloneqq \dotp{~\underbrace{-2\nu_e~\bm{\nu}\bm{\cdot}\nabla^s\bm{v}-q\bm{\nu}}_{\text{symmetry}} + \underbrace{\eta_\Delta\bm{v}}_{\text{penalty}}}{~\underbrace{\bm{u}-\chi_{\Gamma_i}\bm{g}_i}_{\text{b.c.}}}_{\Gamma_e}\quad,\\
\mathscr{N}_\nabla(\bm{v},\bm{u};\bm{g}_i) &\coloneqq -\dotp{\bm{v}}{(\bm{u}-\chi_{\Gamma_i}\bm{g}_i)(\bm{u}\bm{\cdot}\bm{\nu})}_{\Gamma_{\text{in}}} + \dotp{\eta_\nabla~\bm{v}\bm{\cdot}\bm{\nu}}{(\bm{u}-\chi_{\Gamma_i}\bm{g}_i)\bm{\cdot}\bm{\nu}}_{\Gamma_e}\quad,
\end{align}
\end{subequations}
where $\mathscr{N}_\Delta$ is the Stokes problem Nitsche term, and $\mathscr{N}_\nabla$ is the extra Nitsche term needed to impose the b.c. in convection-dominant flows. The characteristic function, $\chi_{\Gamma_i}$, takes the value of $1$ on $\Gamma_i$, and $0$ otherwise, and the inflow boundary, $\Gamma_{\text{in}}$, is such that
\begin{equation}
\Gamma_{\text{in}} \coloneqq \big\{x\in\Gamma_e ~:~ \bm{u}\bm{\cdot}\bm{\nu} < 0 \big\} \label{eq:inflow_bndry}\quad.
\end{equation}
The Nitsche penalty parameters, $\eta_\Delta$ and $\eta_\nabla$, will, in general, scale with the flow, and they will be precisely defined in section \ref{sec:numerics}, where we discuss the discrete form of $\mathscr{M}$. It is also worth noting that the terms $\mathscr{N}_\Delta, \mathscr{N}_\nabla$ vanish when $\bm{u}\equiv\chi_{\Gamma_i}\bm{g}_i$ (consistency), and that the Stokes Nitsche term, $\mathscr{N}_\Delta$, is designed such that the Stokes b.v.p. produces a symmetric weak form. The weak form of the N--S problem, $\mathscr{M}$, is then given by
\begin{equation}
\mathscr{M}(\bm{u},p,\bm{v},q;\bm{x}) \coloneqq \underbrace{a(\bm{v},\bm{u};\bm{x}) + b(p,\bm{v};\bm{x})}_{\text{momentum},~\mathscr{M}_\text{I}}~\underbrace{-~b(q,\bm{u};\bm{x})}_{\textrm{div}\text{-free},~\mathscr{M}_\text{II}}~\underbrace{-j_{\bm{u}}(\bm{v};\bm{x})-j_p(q;\bm{x})}_{\text{inhomogeneous b.c.}} = 0\quad.
\label{eq:ns_weak_form}
\end{equation}
The velocity-velocity bilinear form, $a$, consists of the convective form, $a_{\nabla}$, the viscous form, $a_\Delta$, and the Nitsche form, $a_\mathscr{N}$, which are given by
\begingroup
\allowdisplaybreaks
\begin{subequations}\label{eq:00_bform}
\begin{align}
a(\bm{v},\bm{u};\bm{x}) \coloneqq ~& a_{\nabla} + a_\Delta + a_\mathscr{N} \tag{\ref*{eq:00_bform}}\quad,\\
a_{\nabla}(\bm{v},\bm{u};\bm{x}) \coloneqq ~& \dotp{\bm{v}}{\bm{u}\bm{\cdot}\nabla\bm{u}}_\Omega -\dotp{\bm{v}}{(\bm{u}\bm{\cdot}\bm{\nu})\bm{u}}_{\Gamma_{\text{in}}}\quad,\\
a_\Delta(\bm{v},\bm{u};\bm{x}) \coloneqq ~&\dotp{2\nu_e\nabla^s\bm{v}}{\nabla^s\bm{u}}_{\Omega} - \dotp{2\nu_e\bm{v}}{\bm{\nu}\bm{\cdot}\nabla^s\bm{u}}_{\Gamma_e}-\dotp{2\nu_e\bm{\nu}\bm{\cdot}\nabla^s\bm{v}}{\bm{u}}_{\Gamma_e}\quad, \\
a_\mathscr{N}(\bm{v},\bm{u};\bm{x}) \coloneqq ~& \dotp{\eta_\Delta\bm{v}}{\bm{u}}_{\Gamma_e}
+ \dotp{\eta_\nabla\bm{v}}{(\bm{u}\bm{\cdot}\bm{\nu})\bm{\nu}}_{\Gamma_e}\label{eq:bil_form_a_cutfem}\quad.
\end{align}
\end{subequations}
The velocity-pressure bilinear form, $b$, which encapsulates the divergence-free constraint on $\bm{u}$, is given by
\begin{equation}
b(q,\bm{u};\bm{x}) \coloneqq -\dotp{q}{\nabla\bm{\cdot}\bm{u}}_\Omega + \dotp{q\bm{\nu}}{\bm{u}}_{\Gamma_e}\quad.
\end{equation}
The linear forms $j_{\bm{u}}$, $j_p$, are due to the inhomogeneous inlet and outlet b.c., and are given by
\begin{align}
j_{\bm{u}}(\bm{v};\bm{x}) \coloneqq ~& - \dotp{\bm{v}}{(\bm{u}\bm{\cdot}\bm{\nu})T_{\Gamma_i}\bm{g}_i}_{\Gamma_{\text{in}}} - \dotp{2\nu_e\bm{\nu}\bm{\cdot}\nabla^s\bm{v}}{T_{\Gamma_i}\bm{g}_i}_{\Gamma_i} + \dotp{\eta_\Delta\bm{v}}{T_{\Gamma_i}\bm{g}_i}_{\Gamma_i}\nonumber\\
&+ \dotp{\eta_\nabla\bm{v}}{(T_{\Gamma_i}\bm{g}_i\bm{\cdot}\bm{\nu})\bm{\nu}}_{\Gamma_i} -\dotp{\bm{v}}{T_{\Gamma_o}\bm{g}_o}_{\Gamma_o}\quad, \\
j_p(q;\bm{x}) \coloneqq ~& \dotp{-q\bm{\nu}}{T_{\Gamma_i}\bm{g}_i}_{\Gamma_i}\quad.
\end{align}

\subsubsection{Linearised N--S problem}
The first order Taylor expansion of the weak form of the N--S b.v.p., $\mathscr{M}$, around $\bm{u}_k$, can be written as\footnote{\RREV{The first variation of $\mathscr{M}$ around $\bm{u}_k$ with respect to $\bm{v}$ and $q$ is zero because $\bm{u}_k$ is a N--S solution.}}
\begin{equation}
\bm{y}' = \mathcal{A}_k~\bm{x}' \quad\text{or}\quad\bm{y} \simeq \bm{y}_k + \mathcal{A}_k~\big(\bm{x}-\bm{x}_k\big)\quad,
\label{eq:taylor_state_perturbation}
\end{equation}
where $\bm{y}\equiv(\bm{u},p)$ is the N--S state, $\bm{y}'\equiv(\bm{u}',p')$ is the N--S state perturbation, and $\bm{x}'$ is the N--S parameter perturbation (see also formulas \eqref{eq:def_of_a}, \eqref{eq:z_taylor_exp} in section \ref{sec:nonlinearity_bayes}). The operator $\mathcal{A}_k$ can be split such that
\begin{equation}
\mathcal{A}_k \equiv \big(D^{\bm{y}}_{\bm{x}}\big)_k \coloneqq \big(D^{\mathscr{M}}_{\bm{y}}\big)^{-1}_k~\big(D^{\mathscr{M}}_{\bm{x}}\big)_k\quad.
\label{eq:operator_A_split}
\end{equation}
We therefore proceed to define $(D^{\mathscr{M}}_{\bm{y}})_k$ by linearising $\mathscr{M}$ around $\bm{u}_k$ such that
\begin{alignat}{2}
\delta_{\bm{u}}\mathscr{M} \coloneqq ~& \dotp{\bm{v}}{(D^{\mathscr{M}_\text{I}}_{\bm{u}})_k~\bm{u}'}_{\Omega} + \dotp{q}{(D^{\mathscr{M}_\text{II}}_{\bm{u}})_k~\bm{u}'}_{\Omega} &&= \delta_{\bm{u}}\alpha(\bm{v},\bm{u}') - \delta_{\bm{u}}b(q,\bm{u}')\quad,\\
\delta_{p}\mathscr{M} \coloneqq ~& \dotp{\bm{v}}{(D^{\mathscr{M}_\text{I}}_{p})_k~p'}_\Omega &&=  \delta_pb(p',\bm{v})\quad,
\end{alignat}
where
\begin{subequations}
\begin{equation}
\delta_{\bm{u}} a = \delta_{\bm{u}} a_\nabla + \delta_{\bm{u}} a_\Delta + \delta_{\bm{u}} a_{\mathscr{N}}\quad, 
\end{equation}
\begin{equation}
\delta_{\bm{u}} a_\nabla = \dotp{\bm{v}}{\bm{u}'\bm{\cdot}\nabla\bm{u}_k}_\Omega + \dotp{\bm{v}}{\bm{u}_k\bm{\cdot}\nabla\bm{u}'}_\Omega -\dotp{\bm{v}}{(\bm{u}'\bm{\cdot}\bm{\nu})~\bm{u}_k}_{\Gamma_{\text{in}}} -\dotp{\bm{v}}{(\bm{u}_k\bm{\cdot}\bm{\nu})~\bm{u}'}_{\Gamma_{\text{in}}}\quad,
\end{equation}
\begin{equation}
\delta_{\bm{u}} a_\Delta = a_\nabla(\bm{v},\bm{u}') \quad,\quad \delta_{\bm{u}} a_{\mathscr{N}} = a_{\mathscr{N}}(\bm{v},\bm{u}') \quad,\quad \delta_{\bm{u}} b = b(q,\bm{u}') \quad,\quad \delta_{p} b = b(p',\bm{v})\quad.
\end{equation}
\end{subequations}
The operator $(D^{\mathscr{M}}_{\bm{y}})_k$ is then given by
\begin{gather}
(D^{\mathscr{M}}_{\bm{y}})_k\equiv (D^{\mathscr{M}}_{(\bm{u},p)})_k\coloneqq
\begin{pmatrix}
D^{\mathscr{M}_\text{I}}_{\bm{u}} & D^{\mathscr{M}_\text{I}}_p\\
D^{\mathscr{M}_\text{II}}_{\bm{u}} & 0
\end{pmatrix}_k
\quad.
\label{eq:operator_dmodel_du}
\end{gather}
We proceed to define $\big(D^{\mathscr{M}}_{\bm{x}}\big)_k$ for each N--S parameter, but first we must introduce the adjoint N--S problem. Given that the optimality conditions are based on a linearisation around $\bm{u}_k$, in the following subsections the subscript $(.)_k$ will often be omitted for the sake of clarity.

\subsubsection{Adjoint N--S problem}
For any N--S parameter, $x$, we have
\begin{align}
\delta_x(\mathscr{U}+\mathscr{M}) = \underbrace{\dotp{D_{\bm{u}}\mathscr{U}}{\bm{u}'}_{I_m} + \dotp{\bm{y}^*}{D^{\mathscr{M}}_{\bm{y}}\bm{y}'}_{\Omega}}_{\text{variations that depend on }\bm{y}'} + \dotp{\bm{y}^*}{D^\mathscr{M}_x x'}
\end{align}
where $\bm{u}' = {D}^{\bm{u}}_x~x'$, $p' = {D}^p_x~x'$, and $\bm{y}^*\equiv(\bm{v},q)$. In order to eliminate the dependence of $\delta_x\mathscr{J}$ on state perturbations, $\bm{y}'\equiv(\bm{u}',p')$, we set
\begin{equation}
\dotp{D_{\bm{u}}\mathscr{U}}{\bm{u}'}_{I_m} + \dotp{\bm{y}^*}{D^{\mathscr{M}}_{\bm{y}}\bm{y}'}_{I_m} = 0 \quad.
\end{equation}
In particular, we demand that
\begin{equation}
\Bigg\{\begin{pmatrix}
D_{\bm{u}}\mathscr{U}\\
0
\end{pmatrix}+
\begin{pmatrix}
D^{\mathscr{M}_\text{I}}_{\bm{u}} & D^{\mathscr{M}_\text{I}}_p\\
D^{\mathscr{M}_\text{II}}_{\bm{u}} & 0
\end{pmatrix}^*
\begin{pmatrix}
\bm{v}\\q
\end{pmatrix}\Bigg\}\bm{\cdot} \bm{y}'= 0 \quad \text{for all} \quad \bm{y}' \in \bm{H}^1(\Omega)\times L^2(\Omega)\quad.
\end{equation}
The adjoint state, $\bm{y}^*$, is then obtained by solving the following operator equation
\begin{equation}
A\bm{y}^* = b \quad,\quad \text{where}\quad A \equiv \big(D^{\mathscr{M}}_{\bm{y}}\big)^* \quad,\quad \text{and}\quad b \equiv \begin{pmatrix}-D_{\bm{u}}\mathscr{U}&0\end{pmatrix}^T \quad,
\label{eq:adj_problem_operator_eq}
\end{equation}
which we call the \emph{adjoint N--S problem}, and whose strong form is given by
\begin{gather}
\left\{\begin{alignedat}{2}
-\bm{u}_k\bm{\cdot}\big(2\nabla^s\bm{v}\big) -\nabla\bm{\cdot}\big(2{\nu_e}\nabla^s\bm{v}\big) + \nabla q &= -\big(D_{\bm{u}}\mathscr{U}\big)_k \quad &&\textrm{in}\quad \Omega \\ 
\nabla \bm{\cdot} \bm{v} &= 0 \quad &&\textrm{in}\quad \Omega\\
\bm{v} &= \bm{0} \quad &&\textrm{on}\quad \Gamma\cup\Gamma_i\\
{(\bm{u}_k\bm{\cdot}\bm{\nu})\bm{v}}+{(\bm{u}_k\bm{\cdot}\bm{v})\bm{\nu}}+2{\nu_e}\nabla^s\bm{v}\bm{\cdot}\bm{\nu}-q\bm{\nu} &= \bm{0} \quad &&\textrm{on}\quad \Gamma_o
\end{alignedat}\right. \quad.
\label{eq:navier-stokes_adjoint_problem}
\end{gather}
Note that, $\bm{v} \equiv \bm{0}$ and $q\equiv0$ when $\big(D_{\bm{u}}\mathscr{U}\big)_k \equiv \bm{0}$. Also, on the boundary $\Gamma_e$, the \emph{inhomogeneous} Dirichlet b.c. in the N--S problem produces an \emph{homogeneous} Dirichlet b.c. in the adjoint N--S problem. By solving equation \eqref{eq:adj_problem_operator_eq} for $\bm{y}^*$ we eliminate the dependence of $\delta_x\mathscr{J}$ to state perturbations, $\bm{y}'$. This allows us to define $D^{\mathscr{M}}_{\bm{x}}$ using only the state, $\bm{y}_k$, and the adjoint state, $\bm{y}^*$. We therefore proceed to define $D^{\mathscr{M}}_{x}$ for the N--S parameters $x \in \big(\bm{g}_i,\bm{g}_o,\nu_e,\Omega\big)$.

\subsubsection{Inlet boundary condition \texorpdfstring{$\bm{g}_i$}{}}
The generalised gradient, $D_x\mathscr{M}$, and the operator $D_{x}^\mathscr{M}$ of the \emph{inlet} b.c. are given by
\begin{align}
\dotp{D_{\bm{g}_i}\mathscr{M}}{~\bm{g}'_i}_{I_m} &= \delta_{\bm{g}_i}(-j_{\bm{u}}-j_p) \nonumber\\ &\simeq  \dotp{2\nu_e\bm{\nu}\bm{\cdot}\nabla^s\bm{v}+q\bm{\nu}}{~T_{\Gamma_i}\bm{g}'_i}_{\Gamma_i}\nonumber\\
&=\dotp{T^*_{\Gamma_i}\big(2\nu_e\bm{\nu}\bm{\cdot}\nabla^s\bm{v}+q\bm{\nu}\big)}{~\bm{g}'_i}_{I_m} \label{eq:g_i_grad}\\
&  \equiv \dotp{\big(D_{\bm{g}_i}^\mathscr{M}\big)^*\bm{y}^*}{~\bm{g}'_i}_{I_m} = \dotp{\bm{y}^*}{~D_{\bm{g}_i}^\mathscr{M}\bm{g}'_i}_{I_m} \nonumber\quad.
\end{align}
The approximation in \eqref{eq:g_i_grad} is due to the fact that the adjoint N--S b.c.\ on $\Gamma_e$, $\bm{v}|_{\Gamma_e} \equiv \bm{0}$, is weakly imposed. Boundary terms involving $\bm{v}$ on $\Gamma_e$ can be assumed to be negligible, but not exactly zero (because the penalisation parameters $\eta_\nabla, \eta_\Delta$ are finite).

\subsubsection{Outlet boundary condition \texorpdfstring{$\bm{g}_o$}{}}
For the \emph{outlet} b.c. we find
\begin{align}
\dotp{D_{\bm{g}_o}\mathscr{M}}{~\bm{g}'_o}_{I_m} &= \delta_{\bm{g}_o}(-j_{\bm{u}}) \nonumber\\
&=  \dotp{\bm{v}}{~T_{\Gamma_o}\bm{g}'_o}_{\Gamma_o}\nonumber\\
&=\dotp{T^*_{\Gamma_o}\bm{v}}{~\bm{g}'_o}_{I_m} \label{eq:g_o_grad}\\
& \equiv \dotp{\big(D_{\bm{g}_o}^\mathscr{M}\big)^*\bm{y}^*}{~\bm{g}'_o}_{I_m} = \dotp{\bm{y}^*}{~D_{\bm{g}_o}^\mathscr{M}\bm{g}'_o}_{I_m} \nonumber \quad.
\end{align}

\subsubsection{Effective viscosity field $\nu_e$}
For the \emph{effective} viscosity field we find
\begin{align}
 \dotp{D_{\nu_e}\mathscr{M}}{~\nu'_e}_{\Omega} &=\delta_{\nu_e}a_\Delta \nonumber\\
& \simeq  \dotp{2~\nabla^s\bm{v}\bm{:}\nabla^s\bm{u}}{\nu'_e}_{\Omega} \label{eq:nu_e_grad}\\
&  \equiv \dotp{\big(D_{\nu_e}^\mathscr{M}\big)^*\bm{y}^*}{~\nu_e'}_{\Omega} = \dotp{\bm{y}^*}{~D_{\nu_e}^\mathscr{M}\nu'_e}_{\Omega} \nonumber\quad,
\end{align}
where \RREV{$\bm{a}\bm{:}\bm{b} = \sum_{ij}a_{ij}b_{ij}$}, and the approximation in \eqref{eq:nu_e_grad} is again due to the fact that $\bm{v}|_{\Gamma_e} \equiv \bm{0}$ is weakly imposed, similarly to equation \eqref{eq:g_i_grad}. The generalised gradient of the effective viscosity field, $D_{\nu_e}\mathscr{M}$, which is given by equation \eqref{eq:nu_e_grad}, is, in general, a function in $L^2(\Omega)$. For Newtonian fluids $\nu_e$ is constant (i.e.\ $\nu_e \in \R$) and the generalised gradient is
\begin{equation}
 \dotp{D_{\nu_e}\mathscr{M}}{~\nu'_e}_{\R} = \Dotp{2~\int_\Omega\nabla^s\bm{v}\bm{:}\nabla^s\bm{u}}{~\nu'_e}_{\R}\quad. \label{eq:nu_e_grad_const}
\end{equation}
For `power-law' non-Newtonian fluids, also known as \emph{generalised} Newtonian fluids, the effective viscosity field is an explicit function of the shear-strain magnitude, $\dot{\gamma}\coloneqq\sqrt{2\nabla^s\bm{u}\bm{:}\nabla^s\bm{u}}$, and the $n$ rheological parameters of the fluid, $\bm{p}_{\nu} \in \R^n$, such that
\begin{equation}
\nu_e = \nu_e(\dot{\gamma};\bm{p}_{\nu})\quad.
\label{eq:nu_e_algebraic_model}
\end{equation}
In that case, the N--S unknown is $\bm{p}_{\nu}$ instead of $\nu_e$, and the generalised gradient is thus given by
\begin{equation}
 \dotp{D_{\bm{p}_\nu}\mathscr{M}}{~\bm{p}'_\nu}_{\R^n} = \Dotp{2~\int_\Omega\big(D_{\bm{p}_\nu}\nu_e\big)\big(\nabla^s\bm{v}\bm{:}\nabla^s\bm{u}\big)}{~\bm{p}'_\nu}_{\R^n}\quad,
 \label{eq:nu_e_grad_nonnewtonian}
\end{equation}
where $D_{\bm{p}_{\nu}} \nu_e$ is obtained by differentiating the algebraic relation $\nu_e(\dot{\gamma};\bm{p}_{\nu})$. Note that, when $\nu_e$ is given by the algebraic model \eqref{eq:nu_e_algebraic_model}, which depends on $\dot{\gamma} = \dot{\gamma}(\bm{u})$, $\nu_e$ becomes a function of the velocity field and, consequently, the linearised N--S operator, $(D^{\mathscr{M}}_{\bm{y}})_k$, which is given by \eqref{eq:operator_dmodel_du}, involves additional terms that depend on the variation of $\nu_e$ with respect to $\bm{u}$-perturbations.

It is worth mentioning that viscoelastic non-Newtonian fluids and Reynolds-averaged turbulent flows can be modelled using an effective viscosity field that is implicitly defined by one or more boundary value problems. The present formulation can therefore be extended to such problems by augmenting the objective functional, $\mathscr{J}$, which is given by equation \eqref{eq:obj_func}, with the weak form of the model that governs $\nu_e \in L^2(\Omega)$. This is out of the scope of the present study. Here, we assume that the fluid is Newtonian. Nevertheless, the same framework has been applied to infer the rheological parameters of a power-law non-Newtonian fluid in \cite{Kontogiannis2024b}.

\subsubsection{Geometry $\Omega$}
\label{subsec:geom}
The geometry, $\Omega$, is implicitly defined by the viscous signed distance field (vSDF), $\sdist$, such that
\begin{equation}
\Omega \coloneqq \big\{x\in\Omega~:~ \sdist(x) < 0 \big\}\quad\text{and}\quad \Gamma \coloneqq \big\{x\in\Omega~:~ \sdist(x) = 0 \big\}\quad.
\end{equation}
The vSDF is a \emph{viscosity solution} of an Eikonal problem whose boundary value problem is given by
\begin{equation}
  \left\{\begin{alignedat}{2}
    \sgn(\sdist)~\big(\abs{\nabla\sdist}-1\big) - \eps_\sdist \Delta\sdist &= 0 \quad &&\textrm{in}\quad I_m \\
    \sdist &= 0 \quad &&\textrm{on}\quad \Gamma\\
  \end{alignedat}\right.\quad,
  \label{eq:vsdf_bvp}
\end{equation}
where $\sgn(\bm{\cdot})$ is the sign function, \RREV{$\abs{\cdot}$ is the Euclidean norm}, and $\eps_\sdist$ is the (artificial) viscosity. With the aid of $\sdist$, we can furthermore define the unit normal vector extension, $\nuext$, and the unit normal outward-facing vector extension, $\nuext_o$, such that
\begin{equation}
\nuext \coloneqq \frac{\nabla\sdist}{\abs{\nabla\sdist}} \quad\text{and}\quad \nuext_o \coloneqq \sgn(\sdist)~\nuext\quad.
\label{eq:normal_vec_extension}
\end{equation}
A common approach to modelling geometric (shape) perturbations in shape optimisation \cite{Sokolowski1992,Walker2015}, is by introducing the speed field 
\begin{equation}
\mathscr{V} \coloneqq \zeta\bm{\nu}\quad,\quad \zeta \in {L}^2(\Gamma)\quad,
\end{equation}
where $\zeta$ is the shape perturbation magnitude, and $\bm{\nu}$ is the unit normal vector on $\Gamma$. Then, the vSDF perturbation problem for $\varphi'_\pm$, due to $\Gamma$-perturbations induced by $\mathscr{V}$, is given by
\begin{equation}
  \left\{\begin{alignedat}{2}
    \nuext_o\bm{\cdot}\nabla\varphi'_\pm - \eps_\sdist \Delta\varphi'_\pm &= 0 \quad &&\textrm{in}\quad I_m \\
    \varphi'_\pm +\zeta\partial_{\bm{\nu}}\sdist &=0  \quad &&\textrm{on}\quad \Gamma\\
  \end{alignedat}\right.\quad.
  \label{eq:vsdf_perturb_bvp}
\end{equation}
Problem \eqref{eq:vsdf_perturb_bvp} models the convection-diffusion of $\varphi'_\pm$ with unit velocity, since $\abs{\nuext_o}\equiv1$. Thus it makes sense to define the Reynolds number
\begin{equation}
\Rey_\sdist \coloneqq \ell_\Gamma~\eps_\sdist^{-1}\quad,
\end{equation}
where $\ell_\Gamma$ is a characteristic length scale. \REV{The dissipation of small-scale (geometric) features away from $\Gamma$ can then be explained by the scaling
\begin{equation}
\delta \sim \big(\eps_\sdist \ell_\Gamma\big)^\frac{1}{2} = \mathit{Re}^{-\frac{1}{2}}_\sdist~\ell_\Gamma \quad,
\end{equation}
where $\delta$ is the diffusion width of geometric details at distance $\ell_\Gamma$ from $\Gamma$. The use of artificial dissipation, $\eps_\sdist$, is thus particularly useful in mitigating non-physical, small-scale geometric perturbations, when the position of $\Gamma$ is inferred from noisy velocity fields, $\bm{u}^\star$}.

\begin{figure}[h]
\centering
\includegraphics[width=\textwidth]{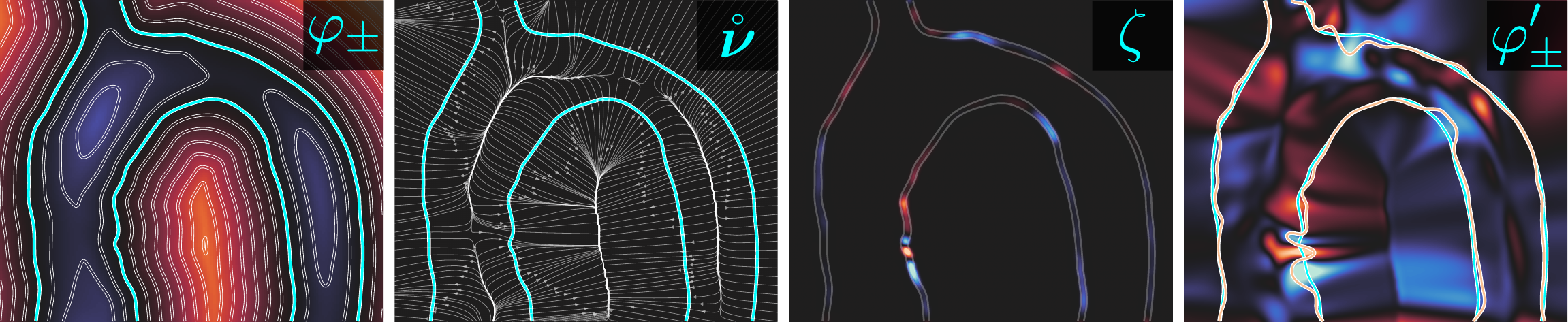}
\caption{The geometry, $\Omega \subset I_m$, is implicitly defined using a viscous signed distance field (vSDF), \mbox{$\sdist \in L^2(I_m)$}, which generates an extension of the unit normal vector on $\partial\Omega$ to the whole domain, $I_m$. The shape gradient, $\zeta \in L^2(\Gamma)$, is then extended to the whole domain, $I_m$, along the $\nuext$-streamlines, by solving a convection-diffusion problem. This produces a perturbation of the vSDF, ${\varphi_\pm'} \in L^2(I_m)$, which models the flow of geometry. The artificial viscosity coefficient, $\eps_\sdist$, controls the regularity of the shape $\partial\Omega$ by dissipating small-scale features when assimilating noisy velocimetry data, $\bm{u}^\star$.}
\label{fig:gflow}
\end{figure}

We now proceed to define the generalised gradient $D_x\mathscr{M}$, and the operator $D_{x}^\mathscr{M}$, for the vSDF. To compute the shape derivatives of the weak form $\mathscr{M}$, due to $\Gamma$-perturbations induced by $\spd$, we use the Leibniz--Reynolds transport theorem (see appendix \ref{app:transp_thm}) to obtain
\begin{align}
 \dotp{D_{\zeta}\mathscr{M}}{~\zeta}_{\Gamma} &\simeq  \dotp{\partial_{\bm{\nu}}\bm{u}_k\bm{\cdot}\big(-2\nu_e~\bm{\nu}\bm{\cdot}\nabla^s\bm{v}-q\bm{\nu}\big)}{\zeta}_{\Gamma} \label{eq:shape_grad}\\
&  \equiv \dotp{\big(D_{\zeta}^\mathscr{M}\big)^*\bm{y}^*}{~\zeta}_{\Gamma} = \dotp{\bm{y}^*}{~D_{\zeta}^\mathscr{M}\zeta}_{\Gamma} \nonumber\quad,
\end{align}
where the approximation in \eqref{eq:shape_grad} is because the homogeneous Dirichlet boundary conditions $\bm{u}_k = \bm{v}=\bm{0}$ are weakly satisfied on $\Gamma$. The weak form of the vSDF b.v.p. \eqref{eq:vsdf_bvp}, $\mathscr{D}$, for test functions $r \in H^1(I_m)$, is given by
\begin{align}
\mathscr{D}(r,\sdist) \coloneqq &\dotp{r}{\sgn(\sdist)~\big(\abs{\nabla\sdist}-1\big)}_{I_m} + \dotp{\eps_\sdist\nabla r }{\nabla\sdist}_{I_m} \nonumber\\ -&\dotp{\eps_\sdist r}{\partial_{\bm{\nu}}\sdist}_{\partial I_m} + \dotp{\eta_\Delta r}{\sdist}_\Gamma = 0 \quad, 
\label{eq:vsdf_weak}
\end{align}
where the homogeneous Dirichlet b.c. is imposed using the penalty term on $\Gamma$. The shape derivatives of the weak form $\mathscr{D}$, due to $\Gamma$-perturbations induced by $\spd$, are then given by
\begin{align}
\delta_\sdist\mathscr{D} = &\dotp{r~\nuext_o}{\nabla\varphi'_{\pm}}_{I_m} + \dotp{\eps_\sdist\nabla r}{\nabla \varphi'_\pm}_{I_m}\nonumber\\ - &\dotp{\eps_\sdist r}{\partial_{\bm{\nu}}\varphi'_\pm}_{\partial I_m} + \dotp{\eta_\Delta r}{\varphi'_\pm}_\Gamma + \dotp{\eta_\Delta r}{\partial_{\bm{\nu}}\sdist~\zeta}_\Gamma \label{eq:vsdf_perturb_weak}\\
\coloneqq& \dotp{r}{D^\mathscr{D}_\sdist ~\varphi'_\pm}_{I_m} + \dotp{r}{D^\mathscr{D}_\zeta ~\zeta}_{\Gamma}\nonumber\quad,
\end{align}
for a given $\zeta \in L^2(\Gamma)$, noting that the weak form \eqref{eq:vsdf_perturb_weak} corresponds to the boundary value problem \eqref{eq:vsdf_perturb_bvp}. We then set
\begin{equation}
\text{i)}~\delta_\sdist\mathscr{D} = 0 \quad \text{for all}\quad r \in L^2(I_m) \quad \text{and} \quad \text{ii)}~\zeta = D_\zeta\mathscr{M} = \partial_{\bm{\nu}}\bm{u}_k\bm{\cdot}\big(-2\nu_e~\bm{\nu}\bm{\cdot}\nabla^s\bm{v}-q\bm{\nu}\big)\quad.
\end{equation}
The generalised gradient $D_\sdist\mathscr{M} \in L^2(I_m)$ is therefore implicitly obtained as the solution of the operator equation
\begin{equation}
A\big(D_\sdist\mathscr{M}\big) = b \quad,\quad \text{where}\quad A \equiv D^{\mathscr{D}}_\sdist \quad,\quad \text{and}\quad b \equiv D^{\mathscr{D}}_\zeta~D_\zeta\mathscr{M} \quad,
\label{eq:sdist_grad}
\end{equation}
which is equivalent to solving \eqref{eq:vsdf_perturb_weak} (or \eqref{eq:vsdf_perturb_bvp} if $\eta_\Delta\to\infty$) for $\zeta = D_\zeta\mathscr{M}$, where $D_\zeta\mathscr{M}$ is given by \eqref{eq:shape_grad}.

\REV{Note that in \cite[Section~2.4]{Kontogiannis2021} the geometry is implicitly represented by an approximate SDF, which is obtained using the `heat method' \cite{Crane2017}, and that this constraint is not incorporated into the objective functional, $\mathscr{J}$. In this paper, instead of the `heat method', we use the viscous Eikonal equation, which is incorporated into $\mathscr{J}$ as an additional constraint. This new constraint is crucial for the solution of the inverse problem in complicated geometries  (e.g. geometries with detailed features and regions of high curvature) because i) the viscous Eikonal equation provides more accurate SDFs, and ii) incorporating this equation as an additional constraint and deriving the respective optimality conditions (see equation \eqref{eq:sdist_grad} for the propagation of the vSDF) ensures numerical consistency and leads to a robust shape inference algorithm.}

\subsubsection{Assembling operator \texorpdfstring{$\mathcal{G}$}{}}
Using the above results we can precisely define $\mathcal{G}^*_k$ (and thus $\mathcal{G}_k$) for the inverse N--S problem. Since $\mathcal{G}_k \coloneqq \mathcal{S}\mathcal{A}_k$, where $\mathcal{S}$ is given by \eqref{eq:model_to_data_proj}, and $\mathcal{A}_k\equiv \big(D^{\bm{y}}_{\bm{x}}\big)_k$ is given by \eqref{eq:operator_A_split}, we find
\begin{equation}
\mathcal{G}^*_k \coloneqq \big(\mathcal{S}D^{\bm{y}}_{\bm{x}}\big)^*_k = \Big(\mathcal{S}\big(D^{\mathscr{M}}_{\bm{y}}\big)^{-1}_k~\big(D^{\mathscr{M}}_{\bm{x}}\big)_k\Big)^* = \big(D^{\mathscr{M}}_{\bm{x}}\big)^*_k~\Big(\big(D^{\mathscr{M}}_{\bm{y}}\big)^*_k\Big)^{-1}\mathcal{S}^*\quad,
\label{eq:def_of_G_adj}
\end{equation}
where $(D^{\mathscr{M}}_{\bm{y}})_k$ is the Jacobian of the N--S problem around $\bm{u}_k$, with respect to the flow variables, which is given by \eqref{eq:operator_dmodel_du}, and 
\begin{equation}
(D^{\mathscr{M}}_{\bm{x}})_k^*\coloneqq\mathrm{diag}\Big(D^{\mathscr{M}}_{\bm{g}_i},D^{\mathscr{M}}_{\bm{g}_o},D^{\mathscr{M}}_{\nu_e},D^{\mathscr{M}}_{\sdist}\Big)_k^*
\label{eq:operator_dmodel_dx}
\end{equation}
is the Jacobian of the N--S problem around $\bm{u}_k$, with respect to the N--S parameters, where $D^{\mathscr{M}}_{\bm{g}_i}$ is given by \eqref{eq:g_i_grad}, $D^{\mathscr{M}}_{\bm{g}_o}$ is given by \eqref{eq:g_o_grad}, $D^{\mathscr{M}}_{\nu_e}$ is given by either \eqref{eq:nu_e_grad}, \eqref{eq:nu_e_grad_const}, or \eqref{eq:nu_e_grad_nonnewtonian}, and $D^{\mathscr{M}}_{\sdist}$ is given by \eqref{eq:sdist_grad}. It is worth noting that $(D^{\mathscr{M}}_{\bm{x}})_k^*$ depends on the flow variables at iteration $k$, $\bm{y}_k$, which are known, and the adjoint flow variables, $\bm{y}^*$, which are found by solving the adjoint problem \eqref{eq:adj_problem_operator_eq}.

\subsubsection{Solving the (forward) N--S problem: operator \texorpdfstring{$\mathcal{Q}$}{}}
The N--S problem \eqref{eq:ns_weak_form} is nonlinear owing to $a_\nabla$ and $\mathscr{N}_\nabla$ being nonlinear. The equivalent Oseen problem, which is linear, is obtained by replacing the nonlinear terms $\bm{u}\bm{\cdot}\nabla\bm{u}$ and $(\bm{u}-\chi_{\Gamma_i}\bm{g}_i)(\bm{u}\bm{\cdot}\bm{\nu})$ with $\bm{\beta}\bm{\cdot}\nabla\bm{u}$ and $(\bm{u}-\chi_{\Gamma_i}\bm{g}_i)(\bm{\beta}\bm{\cdot}\bm{\nu})$, respectively, where $\bm{\beta}$ is a known velocity field. Also, the inflow boundary definition \eqref{eq:inflow_bndry} becomes $\bm{\beta} \bm{\cdot}\bm{\nu} < 0$. The Oseen problem can then be recast as the linear system
\begin{equation}
\label{eq:oseen_problem_system}
\begin{pmatrix}
A_{\bm{\beta}} & B\\
-B^* & 0
\end{pmatrix}
\begin{pmatrix}
\bm{u}\\p
\end{pmatrix}=
\begin{pmatrix}
J_{\bm{u}}\\J_p
\end{pmatrix}\quad,
\end{equation}
where the operators $A_{\bm{\beta}}, B$ and the vectors $J_{\bm{u}},J_p$, correspond to the Oseen linearisation of \eqref{eq:ns_weak_form}.

The N--S problem can be solved using Picard iteration. Starting from an initial guess of the velocity field, $\bm{u}_0$, we solve \eqref{eq:oseen_problem_system} with $\bm{\beta} \mapsfrom \bm{u}_0$ in order to obtain a new velocity field, $\bm{u}_1$. This leads to an iterative process that under certain conditions converges to a fixed point such that $\norm{\bm{u}_{k+1}-\bm{u}_{k}} \to 0$ when $k\to\infty$. The initial guess, $\bm{u}_0$, can be obtained by solving the Stokes problem that corresponds to \eqref{eq:ns_weak_form}. Picard iteration is robust and easy to implement, but converges slowly. In contrast, Newton iteration convergences faster, but requires a good initial approximation of the velocity field. Consequently, to solve N--S we first perform a few Picard iterations in order to obtain a good initial approximation of the velocity field, and then switch to Newton iterations. The Newton iteration is defined through the following update rule
$$\bm{u}_{k+1} \mapsfrom \bm{u}_k + \tau\bm{u}'\quad,\quad p_{k+1} \mapsfrom {p}_k + \tau p'\quad,$$
where $\bm{u}'$, $p'$ are the velocity and pressure perturbations, and $\tau$ is a line search step size. The line search step is initially taken as $\tau =1$ but it can be reduced via a backtracking line search algorithm to facilitate convergence when the initial guess is far away from the solution. The velocity and pressure perturbations are obtained after solving the Newton problem
\begin{equation}
\begin{pmatrix}
D^{\mathscr{M}_\text{I}}_{\bm{u}} & D^{\mathscr{M}_\text{I}}_p\\
D^{\mathscr{M}_\text{II}}_{\bm{u}} & 0
\end{pmatrix}_k
\begin{pmatrix}
\bm{u}'\\\ {p}'
\end{pmatrix}=
\begin{pmatrix}
-R_{\bm{u}_k}\\\ -R_{p_k}
\end{pmatrix}\quad,
\end{equation}
where
\begin{subequations}
\begin{align}
R_{\bm{u}_k} &\coloneqq A_{\bm{u}_k}\bm{u}_k + Bp - J_{\bm{u}}\quad,\\
R_{p_k} &\coloneqq -B^*\bm{u}_k - J_p\quad,
\end{align}
\end{subequations}
are the momentum and mass conservation residuals of the N--S problem, respectively. The Newton problem for the perturbations $\bm{y}'$ can also be written in compact form as $D^\mathscr{M}_{\bm{y}}~\bm{y}' = -\bm{R}$, where $D^\mathscr{M}_{\bm{y}}$ is given by \eqref{eq:operator_dmodel_du}.

To simplify the formulation, and to remain consistent with the operator-based approach we have adopted in this study, we use the nonlinear operator $\mathcal{Q}$, which maps N--S parameters, $\bm{x}$, to N--S velocity and pressure fields, $\bm{y}$, to denote the Picard/Newton iteration approach to solving the N--S problem. When we write
\begin{equation}
\bm{y} = \mathcal{Q}~\bm{x}\quad,
\label{eq:ns_forward_solver}
\end{equation}
we therefore imply that an iterative approach is used to solve the N--S problem for given parameters, $\bm{x}$. Note that, in order to simplify the exposition in section \ref{sec:bayesian_inv}, we defined $\mathcal{Q}$ as the operator that maps parameters, $\bm{x}$, to velocity fields, $\bm{u}$, instead of N--S solutions $(\bm{u},p)$. This abuse of notation, however, does not introduce any problems or inconsistencies in the formulation because $p$ is simply an (internal) Lagrange multiplier that enforces the divergence-free constraint of the N--S problem.

\subsubsection{Solving the viscous Eikonal problem: operator \texorpdfstring{$\mathcal{Q}_{\varphi_\pm}$}{}}
The viscous signed distance field, $\sdist$, is a solution of the viscous Eikonal equation \eqref{eq:vsdf_bvp}, which is nonlinear. We therefore define, similarly to operator $\mathcal{Q}$, an operator $\mathcal{Q}_{\sdist}$ that maps functions $\psi$, whose zeroth level-set is $\partial\Omega$, to viscous signed distance fields (that correspond to $\partial\Omega$), i.e.
\begin{equation}
\sdist = \mathcal{Q}_{\sdist}~\psi\quad.
\label{eq:vsdf_operator}
\end{equation}
Note that $\psi$ encodes the position of the boundary $\Gamma$, which is the input parameter of \eqref{eq:vsdf_bvp}. Equation \eqref{eq:vsdf_operator} implies a standard Newton iteration that solves the weak form of \eqref{eq:vsdf_bvp}, given by \eqref{eq:vsdf_weak}, starting from an approximation of the vSDF, $\psi$. If $\psi$ is not readily available, an initial approximation is obtained by using the `heat method' \cite{Crane2017}\cite[Section~2.4]{Kontogiannis2021}, which involves the solution of two linear and elliptic boundary value problems.

\subsection{N--S parameters prior: operator \texorpdfstring{$\priorcov$}{}}
\label{sec:prior_cov_construction}
The N--S parameter prior terms are given by
\begin{align}
\mathscr{R}_{\bm{x}} \coloneqq \frac{1}{2}\sum_{x \in \bm{x}}\normL{x-\mean{x}}^2_{\mathcal{C}_{\mean{x}}(V_x)}\quad,
\label{eq:bessel_pot}
\end{align}
where $\bm{x} = (\bm{g}_i,\bm{g}_o,\nu_e,\sdist)$, $\mean{x}$ is the prior mean, $V_x$ is the domain, and $\mathcal{C}_{\mean{x}} : L^2(V_x) \to L^2(V_x)$ is the prior covariance operator. For the inlet and outlet boundary conditions, $\bm{g}_i$ and $\bm{g}_o$, we set \cite{Kontogiannis2021}
\begin{equation}
\mathcal{C}_{\mean{x}} = \sigma_x^2~\big(\mathrm{I}-\ell_x^2\widetilde{\Delta}\big)^{-1}\quad,
\end{equation} 
where $\sigma^2_x$ is the prior variance, $\ell_x$ is a characteristic length, and $\widetilde{\Delta}$ is the Laplacian operator \RREV{with zero natural (Neumann) boundary conditions on $\partial V_x$}. For the effective viscosity, $\nu_e$, and the geometry, $\sdist$, we set \mbox{$\mathcal{C}_{\mean{x}} = \sigma^2_x~\mathrm{I}$}. The domain $V_x$ is set to $I_m$ if the unknown is a field, and to $\R^n$, for $n\geq1$, if the unknown is a parameter vector (e.g. viscosity model parameters, or constant viscosity field).

\subsection{Inverse problem solution algorithm}
In section \ref{sec:bayesian_inv} we formulated the general Bayesian inversion problem (MAP estimator) \eqref{eq:map_opt_problem}, which is solved using the update rule \eqref{eq:map_update_rule}, alongside \eqref{eq:map_cov} and \eqref{eq:map_update_gradient}. In the start of section \ref{sec:bayesian_inv_ns_formulation} we formulated the Bayesian inverse Navier--Stokes problem (MAP estimator) using Lagrange multipliers to impose the model constraints. This produced the saddle point problem \eqref{eq:inv_prob_aug}, whose first order optimality conditions were derived in this section. Algorithm \ref{algo_bayesian_inv_ns_problem} summarises the iterative procedure that solves this saddle point problem by collecting the results obtained in section \ref{sec:bayesian_inv_ns_formulation}, and closely follows the nonlinear update rule \eqref{eq:map_update_rule}.
\begin{algorithm}[h]
\small
\textbf{Input:} data, $\bm{u}^\star$, data cov., $\datacov$, prior mean, $\bar{\bm{x}}$, prior cov. $\priorcov$, projection operator, $\mathcal{S}$\\
\textbf{Initialisation:} Solve $\bar{\varphi}_\pm=\mathcal{Q}_{\sdist}~\psi$, $\bar{\bm{y}}=\mathcal{Q}~\bar{\bm{x}}$, and set $\bm{x}_0 \mapsfrom \bar{\bm{x}}$, $\bm{u}_0 \mapsfrom \bar{\bm{u}}$, $\widetilde{\mathcal{C}}_{\bm{x}_0}\mapsfrom\priorcov$, $k\mapsfrom0$\\
\While{$\mathtt{termination\_criterion\_is\_not\_met}$}{
$\makebox[0pt][l]{$\bm{y}^*$}\phantom{\hspace{1.75cm}} \mapsfrom$ \makebox[0pt][l]{$((D^\mathscr{M}_{\bm{y}})_k^*)^{-1}\mathcal{S}^*\datacov^{-1}(\bm{u}^\star-\mathcal{S}\bm{u}_k)\qquad$}\phantom{\hspace{5.5cm}} (solve adjoint N--S problem, eq. \eqref{eq:adj_problem_operator_eq})\\
$\makebox[0pt][l]{$(D_{\bm{x}}\mathscr{M})_k$}\phantom{\hspace{1.75cm}} \mapsfrom$ \makebox[0pt][l]{$(D^\mathscr{M}_{\bm{x}})_k^*~\bm{y}^*$}\phantom{\hspace{5.5cm}} (grad. model term, eq. \eqref{eq:g_i_grad},\eqref{eq:g_o_grad},\eqref{eq:nu_e_grad_const},\eqref{eq:sdist_grad})\\
$\makebox[0pt][l]{$(D_{\bm{x}}\mathscr{J})_k$}\phantom{\hspace{1.75cm}} \mapsfrom$ \makebox[0pt][l]{$(D_{\bm{x}}\mathscr{M}\big)_k + \priorcov^{-1}(\bm{x}_k-\mean{\bm{x}})\qquad$}\phantom{\hspace{5.5cm}} (total gradient, eq. \eqref{eq:map_update_gradient})\\
$\makebox[0pt][l]{$\widetilde{\mathcal{C}}_{\bm{x}_{k+1}}$}\phantom{\hspace{1.75cm}} \mapsfrom$ \makebox[0pt][l]{$(\bm{x}_k,(D_{\bm{x}}\mathscr{J})_k))$}\phantom{\hspace{5.5cm}} (BFGS post. cov. approx., eq. \eqref{eq:bfgs_recursive_formula})\\
$\makebox[0pt][l]{$\bm{x}'$}\phantom{\hspace{1.75cm}} \mapsfrom$ \makebox[0pt][l]{$\widetilde{\mathcal{C}}_{\bm{x}_k} (D_{\bm{x}}\mathscr{J})_k$}\phantom{\hspace{5.5cm}} (parameter perturbation)\\
$\makebox[0pt][l]{$\bm{y}'$}\phantom{\hspace{1.75cm}} \mapsfrom$ \makebox[0pt][l]{$\mathcal{A}_k~\bm{x}'$}\phantom{\hspace{5.5cm}} (solve for state perturbation, eq. \eqref{eq:taylor_state_perturbation})\\
$\makebox[0pt][l]{$\tau$}\phantom{\hspace{1.75cm}} \mapsfrom$ \makebox[0pt][l]{$f(\bm{x}',\bm{y}')$}\phantom{\hspace{5.5cm}} (initial step size)\\
\Do{$\mathscr{J}_k < \mathscr{J}_{k+1}$}{
$\makebox[0pt][l]{$\bm{x}_{k+1}$}\phantom{\hspace{1.1cm}} \mapsfrom$ \makebox[0pt][l]{$\bm{x}_k + \tau\bm{x}'$}\phantom{\hspace{5.5cm}} (update N--S parameters)\\
$\makebox[0pt][l]{$\bm{y}_{k+1}$}\phantom{\hspace{1.1cm}} \mapsfrom$ \makebox[0pt][l]{$\bm{y}_k + \tau\bm{y}'$}\phantom{\hspace{5.5cm}} (update N--S solution)\\
$\makebox[0pt][l]{$\sdist_{k+1}$}\phantom{\hspace{1.1cm}} \mapsfrom$ \makebox[0pt][l]{$\mathcal{Q}_{\sdist}~\psi_{k+1}$}\phantom{\hspace{5.5cm}} (reproject to vSDF space, eq. \eqref{eq:vsdf_operator})\\
$\makebox[0pt][l]{$\bm{y}_{k+1}$}\phantom{\hspace{1.1cm}} \mapsfrom$ \makebox[0pt][l]{$\mathcal{Q}~\bm{x}_{k+1}$}\phantom{\hspace{5.5cm}} (reproject to N--S space, eq. \eqref{eq:ns_forward_solver})\\
$\makebox[0pt][l]{$\tau$}\phantom{\hspace{1.1cm}} \mapsfrom$ \makebox[0pt][l]{$\tau/2$}\phantom{\hspace{5.5cm}} (reduce step size)
}
$\makebox[0pt][l]{$k$}\phantom{\hspace{.25cm}} \mapsfrom$ \makebox[0pt][l]{$k+1$}\phantom{\hspace{5.5cm}}\\
}
\textbf{Output:} parameters, $\bm{x}^\circ \mapsfrom \bm{x}_k$, $\widetilde{\mathcal{C}}_{\bm{x}^\circ}\mapsfrom \widetilde{\mathcal{C}}_{\bm{x}_k}$, and state, $\bm{y}^\circ \mapsfrom \bm{y}_k$ (\emph{MAP estimates})
\caption{Bayesian inversion of the Navier--Stokes problem}\label{algo_bayesian_inv_ns_problem}
\label{algo:bayesian_inversion_ns}
\end{algorithm}

It is worth noting that in order to compute the model term of the gradient, $\mathcal{G}^*_k\mathcal{S}^*\datacov^{-1}(\bm{u}^\star-\mathcal{S}\bm{u}_k)$, which appears in \eqref{eq:map_update_gradient}, we use the adjoint state, $\bm{y}^*$, as an auxiliary variable. This is because, if the algorithm is to be discretised and solved by a computer, the operator $\mathcal{G}^*_k$, as defined by \eqref{eq:def_of_G_adj}, is constructed using the inverse of $(D^\mathscr{M}_{\bm{y}})_k^*$, which is not necessarily sparse and thus prohibitively expensive (or impossible for large-scale problems) to store in computer memory, even with state-of-the-art hardware. The operator $(D^\mathscr{M}_{\bm{y}})_k^*$, however, is known to be sparse and can thus be stored in computer memory. Once the total gradient is computed, gradient information is used to update the posterior parameter covariance approximation, $\widetilde{\mathcal{C}}_{\bm{x}_{k+1}}$, using the BFGS quasi-Newton method. To accelerate the algorithm we do not impose the Wolfe conditions \cite{Fletcher2000} (since they require additional evaluations of the N--S and the adjoint N--S problems). We, however, use damping \cite{Goldfarb2020} to ensure $\postcovxk$ remains positive definite, and its approximation remains numerically stable. The initial step size of the backtracking line search, $\tau$, is obtained by $f(\bm{x}',\bm{y}')$, which is a problem-dependent user-input functional that depends on the parameter and state perturbations norms. During the line search, the perturbed parameters and the state are reprojected to their corresponding constraints (i.e. $\psi_{k+1}\equiv(\sdist)_{k+1}$ must be a vSDF and $\bm{y}$ must be a N--S solution), and the new value of the objective functional, $\mathscr{J}_{k+1}$, is computed. The algorithm terminates if one of the following criteria is met: i) the optimality condition is satisfied within a tolerance $\eps$, i.e. $\delta_{\bm{x}}\mathscr{J} \leq \eps$, ii) $\mathscr{J}$ reduces below a specified tolerance $\eps$, i.e. $\mathscr{J}\leq \eps$, iii) the line search fails to reduce $\mathscr{J}$ further. 

The output of algorithm \ref{algo_bayesian_inv_ns_problem} is the posterior (MAP) parameter mean, $\bm{x}^\circ$, the approximated posterior parameter covariance, $\widetilde{\mathcal{C}}_{\bm{x}^\circ}$, and the posterior (MAP) state mean $\bm{y}^\circ$. The approximated posterior \emph{state} covariance, $\widetilde{\mathcal{C}}_{\bm{y}^\circ}$, is given by \eqref{eq:state_covariance}. We thus obtain the following approximations for the posterior probability density functions
\begin{gather}
\pi(\bm{x}|\bm{u}^\star) \simeq \mathcal{N}(\bm{x}^\circ,\widetilde{\mathcal{C}}_{\bm{x}^\circ})\quad,\quad \pi(\bm{y}|\bm{x}) \simeq \mathcal{N}(\bm{y}^\circ,\widetilde{\mathcal{C}}_{\bm{y}^\circ})\quad.
\end{gather}
The Gaussian approximation of the posterior distribution was discussed in section \ref{sec:laplace_approx}, and uncertainty estimation from covariance operators was discussed in section \ref{sec:uncertainty_estimation}. Note that the accuracy of the computed uncertainties relies on i) the accuracy of the BFGS approximation of ${\mathcal{C}}_{\bm{x}^\circ}$, and ii) the accuracy of the Laplace approximation \eqref{eq:laplace_approx}.

\section{Discretised operators and numerics}
\label{sec:numerics}

We numerically solve the boundary value problems of section \ref{sec:bayesian_inv_ns_formulation} using a `meshless' finite element method (FEM) \cite{Babuska2003}. In particular, we implement the fictitious domain cut-cell finite element method, introduced by \cite{Burman2010,Burman2012} for the Poisson problem, and later on extended to the Stokes and the Oseen problems \cite{Schott2014,Massing2014,Burman2015,Massing2018}. This method has certain advantages over traditional mesh-dependent methods when working with complicated, moving geometries, since it does not require a body fitted mesh. Rather, it uses a fixed Cartesian mesh and thus requires special handling of the cut-cell region. We ensure consistency between the continuous and discrete formulations of the inverse problem \eqref{eq:inv_prob_aug} by discretising the weak forms derived in section \ref{sec:bayesian_inv_ns_formulation} and adding \emph{symmetric} numerical stabilisation terms that vanish as the background mesh size tends to zero.  The numerical stabilisation that is introduced, and the weak imposition of the boundary conditions using Nitsche terms, lead to an adjoint-consistent formulation, i.e. the discretised continuous adjoint operator is equivalent to the discrete adjoint operator. 

\subsection{Stabilised cut-cell finite element method}
\label{sec:stab_cutcell_fem}
In the context of cut-cell FEM, we consider a geometry of interest $\Omega$, immersed in a global domain $I_m\subset\mathbb{R}^3$. The geometry $\Omega$, and its boundary, $\Gamma$, are implicitly defined by a viscous signed distance field, $\sdist \in L^2(I_m)$. We define $\mathcal{T}_h$ to be a tessellation of $I_m$ produced by cuboid cells (voxels), $K \in \mathcal{T}_h$, having sides of length $h$. This tessellation is comprised of cut-cells, $\mathcal{T}_h^\triangleright$, which are cells that are cut by the boundary, $\Gamma$, and of intact cells,  $\mathcal{T}_h^{\smallsquare}$, which are found either inside or outside of $\Omega$, and are not cut by the boundary,  $\Gamma$ (see figure \ref{fig:flow_reconstruction_concept}). We assume that the boundary, $\Gamma$, is well-resolved, i.e. $\ell/h \gg 1$, where $\ell$ is the smallest length scale of $\Gamma$. Because of this assumption, the intersection $\mathcal{T}_h^\triangleright\equiv\mathcal{T}_h \cap \Gamma$ is comprised of specific types of polyhedral elements, $\mathcal{P}$, which are shown in figure \ref{fig:cuboid_plane_intersection2}. The discretised model space, $M_h$, is generated by assigning a trilinear finite element, $\mathcal{Q}_1$, to every cell $K \in \mathcal{T}_h$, i.e.
\begin{gather}
M_h = \Big\{u \in \mathscr{C}^0(I_m)~:~  u\vert_K \in \mathcal{Q}_1 ~\text{for all}~ K \in \mathcal{T}_h \Big\}\quad,
\label{eq:discrete_fem_space}
\end{gather} 
where $\mathscr{C}^0$ is the space of continuous functions, and 
\begin{gather}
\mathcal{Q}_1 \coloneqq \Big\{\sum_\ell c_\ell~ p_\ell(x)q_\ell(y)r_\ell(z), ~p_\ell, q_\ell, r_\ell~\text{polynomials of deg.}\leq 1,~ c_\ell \in \R \Big\}\quad.
\end{gather}
\RREV{The discrete velocity-pressure pair is defined in the same space, i.e. $(\bm{u}_h,p_h) \in \bm{M}_h\times M_h$, which leads to an inf-sup unstable formulation \cite{BrennerSusanneScott2008}. We therefore use the continuous interior penalty (CIP) method for pressure and velocity (convective) stabilisation \cite{Burman2006}, and \mbox{$\nabla$-div} stabilisation for both stabilisation and preconditioning of the Schur complement \cite{Benzi2006,olshanskii_grad-div_2009,Heister2013}. The behaviour of the numerical solution in the vicinity of the intersection $\mathcal{T}_h\cap\Gamma$ is stabilised using a CIP-like approach, known as the ghost-penalty method \cite{Burman2010,Burman2012,Massing2018}.}

\begin{figure}
\centering 
\begin{subfigure}{.76\textwidth}
\centering
  \reflectbox{\includegraphics[height=0.68\linewidth]{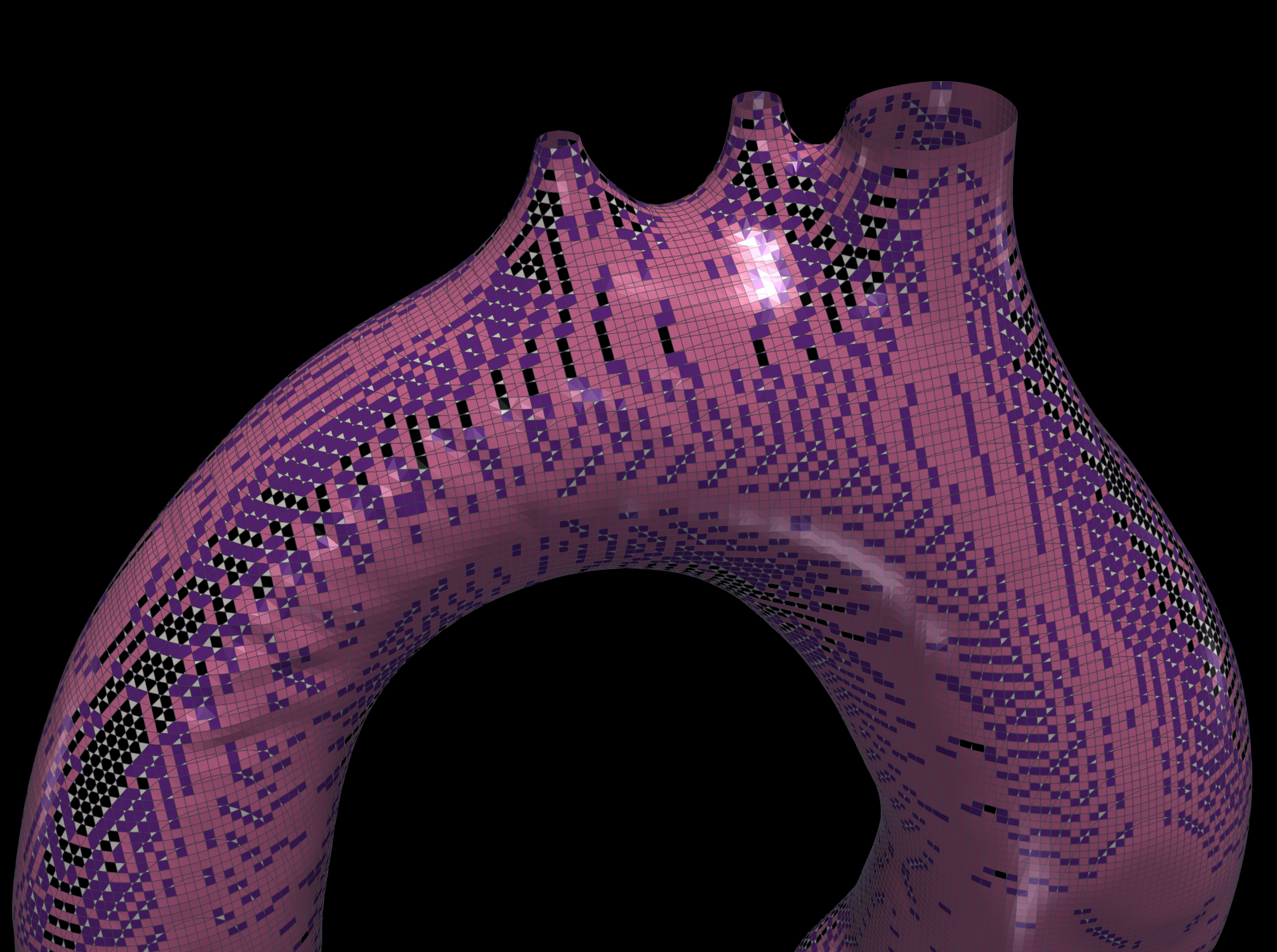}}
  \caption{Cut-cell facets forming the discretised boundary $\Gamma_h$.}
  \label{fig:cut_facets_forming_boundary}
\end{subfigure}%
\begin{subfigure}{.306\textwidth}
\centering
  \includegraphics[height=1.69\linewidth]{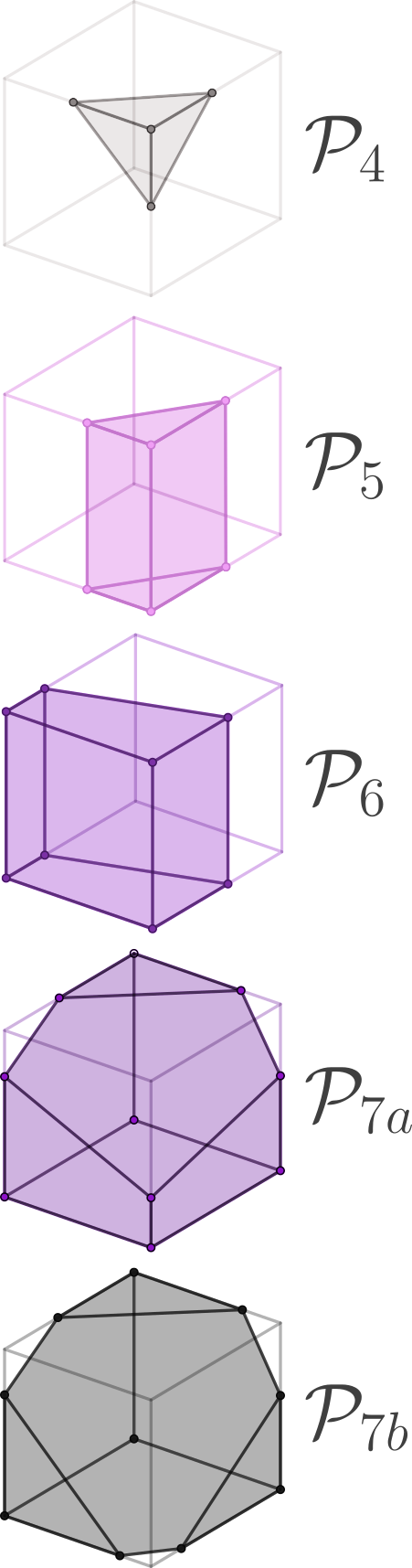}
  \caption{Cut-cell types}
  \label{fig:cuboid_plane_intersection2}
\end{subfigure}%
\caption{In cut-cell FEM the boundary, $\Gamma$, is implicitly defined by the zeroth level-set of $\sdist$. The discretised boundary, $\Gamma_h$, consists of the facets of the cut-cells. Under the assumption of a geometry-resolving mesh, there are only a few different cut-cell types (polyhedra, $\mathcal{P}$, formed by cuboid-plane intersections).}
\end{figure}

\subsubsection{Discrete weak form of the Navier--Stokes problem}
\label{sec:discrete_weak_ns_formulation}
The \RREV{stabilised}, discrete weak form of the N--S problem, $\mathscr{M}_h$, is given by
\begin{gather}
\mathscr{M}_h \coloneqq \mathscr{M}(\bm{u}_h,p_h,\bm{v}_h,q_h;\bm{x}) + \underbrace{s^{\nabla\text{div}}_{{u}_h}(\bm{v}_h,\bm{u}_h) + s^{\text{CIP}}_{{u}_h}(\bm{v}_h,\bm{u}_h) + s^{\text{CIP}}_{p_h}(q_h,p_h)}_{\text{numerical stabilisation terms}} = 0\quad,
\label{eq:ns_discrete_problem}
\end{gather}
where $\mathscr{M}$ is given by \eqref{eq:ns_weak_form}, and the numerical stabilisation terms are given by
\begin{subequations}
\begin{align}
s^{\nabla\text{div}}_{{u}_h}(\bm{v}_h,\bm{u}_h) \coloneqq &\sum_{\mask{F\in \mathcal{F}_i}{K\in \mathcal{T}_h\cap\Omega}} {{c_{\nabla\text{div},K}}}\ \dotp{\nabla\bm{\cdot}\bm{v}_h}{\nabla\bm{\cdot}\bm{u}_h}_{K\cap\Omega}\quad, \\
s^{\text{CIP}}_{{u}_h}(\bm{v}_h,\bm{u}_h) \coloneqq &\sum_{F\in \mathcal{F}_i} \mask{c_{\nabla\text{div},K}}{c_{{u},F}}\ \dotp{[\nabla \bm{v}_h]}{[\nabla \bm{u}_h]}_F\quad, \\
s^{\text{CIP}}_{p_h}(q_h,p_h) \coloneqq & \sum_{F\in \mathcal{F}_i} \mask{c_{\nabla\text{div},K}}{c_{p,F}}\ \dotp{[\nabla q_h]}{[\nabla p_h]}_F \quad.\label{eq:cip_press_stab}
\end{align}
\end{subequations}
where $\mathcal{F}_i$ is the set of interior facets, and $[\cdot]$ is the jump of a function along the face $F$, as defined in \cite{Massing2018}. \REVV{For the numerical method to be robust in a variety of flow conditions, and {for the discrete, linearised N--S solution to be unique \cite{Massing2018}}, the Nitsche penalty parameters, $\eta_\Delta$, $\eta_\nabla$, and the stabilisation coefficients, $c_{\nabla\text{div},K}$, $c_{{u},F}$, and $c_{p,F}$, must all scale with the flow.} Appropriate voxel-wise constant formulas are given by \cite{Codina2002,Massing2018}
\begin{subequations}
\begin{gather}
\eta_\Delta = \gamma_\mathscr{N}~h^{-1}\norm{\nu_e}_{L^\infty(K)}\quad,\quad\eta_\nabla = \gamma_\mathscr{N}~h^{-1}\phi_{u,K}\quad,
\end{gather}
\begin{gather}
c_{\nabla\text{div},K} \coloneqq {\gamma_{\nabla\text{div}}}~\big(\norm{\nu_e}_{L^\infty(K)} + h\norm{\bm{u}_k}_{L^\infty(K)}\big)\quad,
\end{gather}
\begin{gather}
c_{u,F} \coloneqq \left\{
\begin{array}{ll}
       c_{u,F_i}  &, ~F \in \mathcal{F}_i\setminus\mathcal{F}_\Gamma \\
       {\gamma_{\nu}}~h\norm{\nu_e}_{L^\infty(K)} + {c_{{u},F_i}} &, ~F \in \mathcal{F}_\Gamma\\
\end{array} 
\right. \quad,\quad c_{u,F_i} \coloneqq {\gamma_{{u}}}~h\norm{\bm{u}_k}_{L^\infty(F)}^2~\langle\phi_p\rangle\lvert_F\quad, 
\end{gather}
\begin{gather}
 \phi_{p,K} \coloneqq {h^2\phi_{{u},K}^{-1}}\quad,\quad\phi_{{u},K} \coloneqq \norm{\nu_e}_{L^\infty(K)} + {c_{{u}}}h\norm{\bm{u}_k}_{L^\infty(K)} \quad,\quad {c_{p,F}} \coloneqq {\gamma_{p}}~h\langle\phi_p\rangle\lvert_F\quad,
\end{gather}
\end{subequations}
\endgroup
where $\bm{u}_k$ is the velocity field at iteration $k$, $\mathcal{F}_\Gamma \subset \mathcal{F}_i$ is the set of boundary facets, and $\langle\cdot\rangle\lvert_F$ is the face average value between two neighbouring voxels \cite{Massing2018}. Typical values for the numerical parameters are
\begin{gather}
\begin{array}{c c c c c c}
\gamma_\mathscr{N} & \gamma_{\nabla\text{div}} & \gamma_{u} & \gamma_{p} &
\gamma_\nu & c_{u}\\
(30,100) & (0.1,1) & (0.01,0.1) & (0.01,0.1) & (0.01,0.1) & (0.1,1)
\end{array}\quad.
\end{gather}
The numerical (Galerkin) problem is thus to find $(\bm{u}_h,p_h)\in \bm{M}_h\times M_h$ such that \eqref{eq:ns_discrete_problem} holds true for all \mbox{$(\bm{v}_h,q_h)\in \bm{M}_h\times M_h$}. The stabilised cut-cell finite element method is adopted from \cite{Massing2018}. In this study we slightly modify the formulation by including $\nabla$-div stabilisation in order to improve the condition number of the Schur complement \cite{Benzi2006,olshanskii_grad-div_2009,Heister2013}.

\subsubsection{Discrete weak form of the viscous Eikonal problem}
The discrete weak form of the viscous Eikonal problem, $\mathscr{D}_h$, is given by
\begin{gather}
\mathscr{D}_h \coloneqq \mathscr{D}(r_h,\sdist_h) + s^{\text{CIP}}_{u_h}(r_h,\sdist_h) = 0\quad,
\label{eq:vsdf_weak_discrete}
\end{gather}
where $\mathscr{D}$ is given by \eqref{eq:vsdf_weak}, and the CIP stabilisation term, $s^{\text{CIP}}_{u_h}(\cdot,\cdot)$, was defined in section \ref{sec:discrete_weak_ns_formulation}. We further define the smoothed sign function \begin{equation}
\mathrm{sgn}_h(\sdist_h) \coloneqq \frac{\sdist_h}{\sqrt{{\varphi^2_\pm}_h+h^2}}\quad,
\end{equation}
which replaces the sign function, $\mathrm{sgn}(x)\coloneqq x/\abs{x}$, of \eqref{eq:vsdf_weak} in the case of the discrete problem. As in section \ref{sec:discrete_weak_ns_formulation}, the numerical (Galerkin) problem is to find $\sdist_h\in M_h$ such that \eqref{eq:vsdf_weak_discrete} holds true for all \mbox{$r_h\in M_h$}.
\subsection{Cut-cell integration}
\label{sec:cut_cell_integrals}

For the numerical evaluation of integrals we use standard Gaussian quadrature for intact cells, $K \in \mathcal{T}_h^{\smallsquare}$. For cut-cells, $K \in \mathcal{T}_h^\triangleright$, however, integration must be considered only on the intersection $K\cap\Omega$ for volumetric integrals and on $K\cap\Gamma$ for surface integrals. We therefore generalise the approach of \cite{Mirtich1996}, which relies on the divergence theorem and replaces the integral over $K \cap \Omega$ with an integral over $\partial\big(K\cap\Omega\big)$. The boundary integral on $\partial\big(K\cap\Omega\big)$ is then easily computed using one-dimensional Gaussian quadrature \cite{Massing2013}. For instance, let us consider that we want to compute the integral of the Laplacian in a cut-voxel, $\mathcal{P}$, given by
\begin{gather}
A_{pq} \coloneqq \int_\mathcal{P} \nabla\varphi_p \bm{\cdot} \nabla\varphi_q \quad.
\end{gather}
If $\varphi$ is a polynomial test function, we can find coefficients ${\color{black}c_{pqk}}$ such that 
\begin{gather}
A_{pq} = \sum_k {\color{black}c_{pqk}} \int_\mathcal{P} x^{\alpha^k}\quad,
\label{eq:monomial_form_of_integral}
\end{gather}
for a multi-index $\alpha^k$. In general, any integral of the weak (Galerkin) formulation can be expressed in the form of \eqref{eq:monomial_form_of_integral}. The problem is then to i) find the coefficients $c_{pqk}$, and ii) compute the monomial integrals
\begin{gather}
I^\mathcal{P}_k \coloneqq \int_\mathcal{P} x^{\alpha^k}\quad.
\end{gather}
For polynomial test functions, the coefficients $c_{pqk}$ are easily obtained using symbolic mathematics \cite{10.7717/peerj-cs.103}. They only depend on the integrand, which depends on the boundary value problem at hand, and can thus be stored in a computer file. The monomial integrals, $I^\mathcal{P}_k$, depend on the geometry, $\mathcal{P}$, and therefore need to be computed as a pre-processing step before assembling the FEM system matrix. For every cut-cell, the monomial integrals that are needed can be computed only once and stored in memory in order to avoid unnecessary extra calculations. The monomial integrals are evaluated only for the cut-cells,  $K \in \mathcal{T}_h^\triangleright$. Hence, the numerical cost is negligible compared to other bottlenecks in the algorithm, such as the linear system solution. The algorithm to compute $I^\mathcal{P}_k$, which generalises the methodology presented originally in \cite{Mirtich1996}, is described in \cite[Appendix~D]{kontogiannis_2023}.

\subsection{Exact pressure projection via the Schur complement}
\label{sec:schur-complement}
The  ($\bm{u}_h$,$p_h$)-system that results after assembling the discrete weak form \eqref{eq:ns_weak_form}, is given by
\begin{gather}
\begin{pmatrix}
A & B\\
-B^* & D
\end{pmatrix} 
\begin{pmatrix}
\bm{u}_h \\
p_h
\end{pmatrix}
=
\begin{pmatrix}
\bm{f} \\
g
\end{pmatrix}\quad,
\label{eq:vel-pres_system}
\end{gather}
and has a non-zero pressure-pressure block, i.e. $D\neq0$ (in contrast to \eqref{eq:oseen_problem_system})  due to the pressure stabilisation term, $s^{\text{CIP}}_{p_h}(q_h,p_h)$, which has been added in the discrete weak form. 
Multiplying the first equation of \eqref{eq:vel-pres_system} by $A^{-1}$, and substituting it into the second equation, we obtain
\begin{gather}
Sp_h = -B^*A^{-1}\bm{f}-g\quad,
\label{eq:press_proj}
\end{gather}
where $S\equiv\big(-B^*A^{-1}B-D\big)$ is the \emph{Schur complement}. After solving \eqref{eq:press_proj} for $p_h$, $\bm{u}_h$ can be obtained by solving
\begin{gather}
{A\bm{u}_h = \bm{f} - Bp_h} \quad.
\end{gather}
Equivalently, we can write the coupled system \eqref{eq:vel-pres_system} in its decoupled form
\begin{gather}
\begin{pmatrix}
A & 0\\0 & S
\end{pmatrix}
\begin{pmatrix}
\bm{u}_h\\p_h
\end{pmatrix}
=
\begin{pmatrix}
\bm{f} - BS^{-1}\big(-B^*A^{-1}\bm{f}-g\big) \\
-B^*A^{-1}\bm{f}-g
\end{pmatrix}\quad.
\end{gather}
It is worth noting that $S$ is not necessarily sparse, and thus it cannot be assembled. Instead, $S$ is treated as a matrix-vector product operator. This implies that, when an iterative algorithm is used for \eqref{eq:press_proj}, a linear system with the velocity-velocity subblock, $A$, is solved at every pressure iteration. For the exact projection method to be efficient we therefore require good preconditioners for both $S$ and $A$. We also use $\nabla$-div stabilisation for Schur-complement preconditioning, which amounts to adding $s^{\nabla\text{div}}_{{u}_h}(\bm{v}_h,\bm{u}_h)$ to the discrete weak form \cite{Heister2013}. We then use the approximation of $S$,
\begin{gather}
\widetilde{S} \coloneqq -B^*\big(\textrm{diag}(A)\big)^{-1}B-D\quad,
\end{gather}
as the preconditioner of the iterative pressure solver. 

\subsection{Implementation}
The Bayesian inverse Navier--Stokes solver, which incorporates the cut-cell FEM solver, is implemented in Python using MPI (\texttt{mpi4py}, \cite{DALCIN20051108}) and PETSc (\texttt{petsc4py}) \cite{petsc-web-page,petsc-user-ref}. We solve \eqref{eq:press_proj} using FGMRES \cite{doi:10.1137/0914028} for the outer iterations (pressure iterations), and GMRES \cite{doi:10.1137/0907058} for the inner iterations (velocity subblock, $A$, iterations). We precondition the inner solver using either the overlapping additive Schwarz method (ASM) \cite{15bcc4ae196041f1a0457f4a82ea9808}, with ILU(0) for each subdomain, or generalised algebraic multigrid (GAMG). In our implementation, ASM is more efficient when run on a workstation, and for relatively small problems, while GAMG performs and scales better in high performance computing clusters.  When solving with $\widetilde{S}$, we use GMRES with GAMG. A similar setup is used to solve the viscous Eikonal problem. The uncertainties around the MAP solution can be estimated by computing the eigendecomposition of either the exact or the approximated posterior covariance operators, as was described in section \ref{sec:uncert_estim}. This requires the integration of a large-scale eigenvalue problem solver, such as SLEPc \cite{Hernandez:2005:SSF} (\texttt{slepc4py}), into the current software, which is the subject of future work. 

\section{Reconstruction of flow-MRI data in an aortic arch}
\begin{figure}
\centering
\begin{subfigure}{0.645\textwidth}
\centering
\includegraphics[width=0.95\textwidth,trim=5 50 25 45,clip]{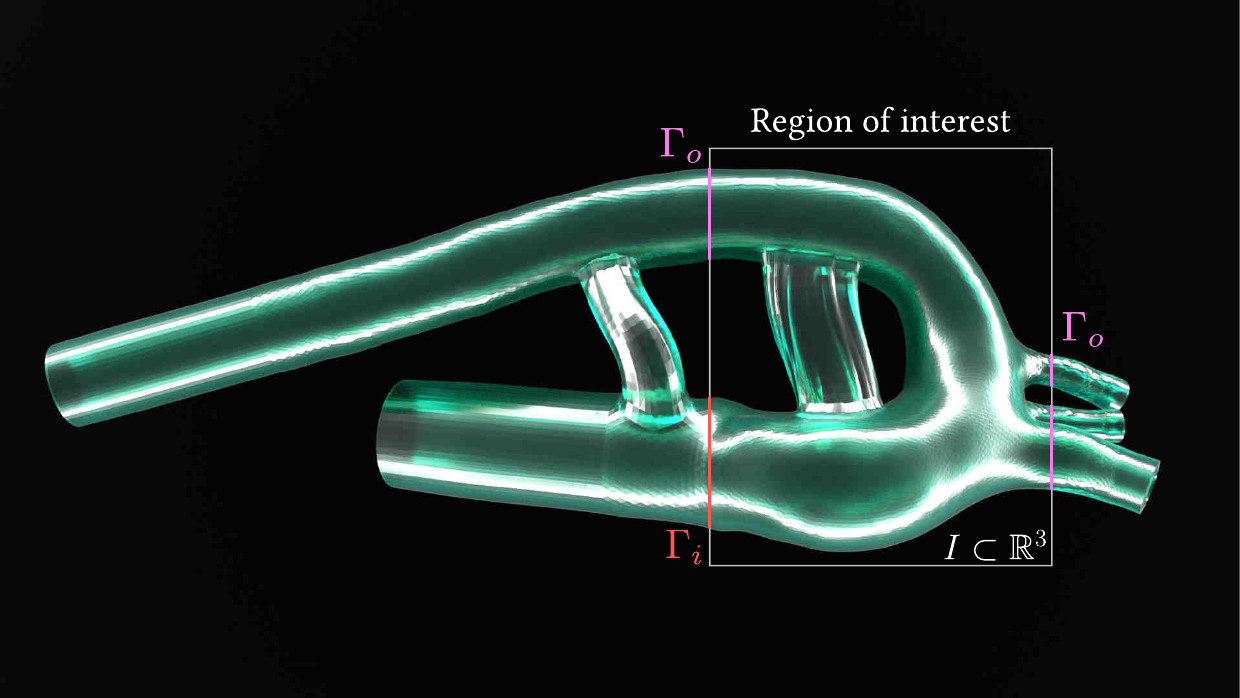}
\caption{Physical model and b.c. setup}
\label{fig:3d_printed_aorta}
\end{subfigure}
\begin{subfigure}{0.325\textwidth}
\centering
\includegraphics[width=0.95\textwidth]{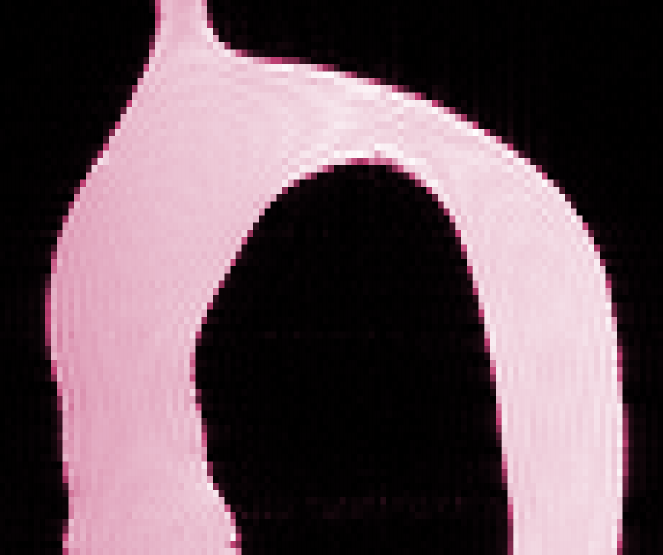}
\caption{spin density (2D slice of ROI)}
\label{fig:spin_density_scan}
\end{subfigure}
\caption{3D-printed (rigid) physical model of an aortic arch and boundary condition setup (figure \ref{fig:3d_printed_aorta}). We have obtained flow-MRI data of the full geometry, but we only reconstruct the flow in the region of interest (ROI), $I$, which is enclosed in a cuboid in $\R^3$. To reconstruct the flow in $I$, we only use the flow-MRI data inside $I$. That is, $I_d \equiv I_m \equiv I$.}
\label{fig:problem_geometry}
\end{figure}

In this section we demonstrate algorithm \ref{algo:bayesian_inversion_ns} by reconstructing flow-MRI data of a steady laminar flow through a physical model of an aortic arch at two different flow conditions: i) low $\Rey$ and ii) high $\Rey$, and for two different signal-to-noise ratios (SNR): i) low SNR ($\sim5$) and ii) high SNR ($>200$).

\subsection{Problem setup}
The model geometry, which is shown in figure \ref{fig:3d_printed_aorta}, is taken from a computed tomography scan and was \mbox{3D-printed} using stereolithography (resin-printing). The characteristic length, based on the inlet diameter, is \mbox{$L=1.3$ cm}, and the characteristic velocity is defined as \mbox{$U\equiv Q/A$}, where $Q$ is the volumetric flow rate, and $A$ is the cross-sectional area of the inlet, $\Gamma_i$. In the low $\Rey$ case we have $\Rey = 554$ ($= UL/\nu$), where $U = 4.26$ cm/s, and in the high $\Rey$ case we have $\Rey = 1526$ ($= UL/\nu$), where $U = 11.74$ cm/s. The field of view (FOV) is $5.39\times2.53\times4.51$ cm, which is the region of interest (ROI) (shown in figure \ref{fig:3d_printed_aorta}), and is the same for both the data space and the model space. That is, we reconstruct the flow in the ROI using only the flow-MRI data inside the ROI. The number of voxels and the resolution in the data space and the model space are shown in table \ref{tab:input_params_aorta_3d}. Note that, the total number of \emph{active} voxels in the model space are $\sim 670\times10^3$, consisting of $\sim 590\times10^3$ interior intact voxels and $\sim 80\times10^3$ cut voxels. For this geometry we have obtained quantitatively accurate high SNR ($>200$) flow-MRI (phase-contrast MRI) data, which we corrupt with Gaussian white noise in order to generate the low SNR ($\sim5$) \mbox{flow-MRI} data. (Further details regarding the flow-MRI experiment can be found in \cite[Appendix~E.2]{kontogiannis_2023}, and the SNR is computed as in \cite[Section~III.A]{Kontogiannis2022b}.) The noise magnitude of the low SNR and the high SNR data is shown in table \ref{tab:input_params_aorta_3d}. In the \mbox{N--S} problem \eqref{eq:navierstokes_bvp}, we set a no-slip (zero-Dirichlet) b.c. on the aorta walls, $\Gamma$, a Dirichlet (velocity) boundary condition, $\bm{g}_i$, on the ascending aorta inlet, $\Gamma_i$, and natural boundary conditions, $\bm{g}_o$, on the aorta outlets, $\Gamma_o$, which are four in total, consisting of the three upper branches and the descending aorta (shown in figure \ref{fig:3d_printed_aorta}).

\begin{table}
\small
  \begin{center}
\def~{\hphantom{0}}
\vspace{0.35cm}
\centering
  \begin{tabular}{ccc|ccc}
        \multicolumn{6}{c}{\emph{data and model domain}}\\[3pt]\hline\\[-1.0em]
        \multicolumn{2}{c}{data voxels}   &  {data resolution} &  \multicolumn{2}{c}{model voxels} & {model resolution}\\[3pt]
        \multicolumn{2}{c}{$115\times54\times77$} & {$469\times469\times586$ $\mu$m}  & \multicolumn{2}{c}{$207\times97\times174$} & {$260\times260\times260$ $\mu$m}\\[6pt] 
  \end{tabular}\\
  \centering
  \begin{tabular}{cccc}
           \multicolumn{4}{c}{\emph{data noise level $(\sigma_{u^\star_x},\sigma_{u^\star_y},\sigma_{u^\star_z})/U$}} \\[3pt]\hline\\[-1.0em]
          & low SNR & high SNR & $U$ (cm/s) \\[3pt]
          low $\Rey$ & $(0.519,0.519,0.519)$ & $(0.011,0.010,0.011)$ & $4.26$ \\[3pt]
          high $\Rey$ & $(0.544,0.544,0.545)$& $(0.012,0.012,0.020)$ & $11.74$\\[6pt]    
  \end{tabular}\\
  \centering
  \begin{tabular}{cccccccc}
           \multicolumn{8}{c}{\emph{prior covariance ($\priorcov$) and vSDF parameters}} \\[3pt]\hline\\[-1.0em]
          $\sigma_\sdist$ & $\sigma_{\bm{g}_i}/U$ & $\sigma_{\bm{g}_o}L/U$ & $\sigma_\nu$ & $\ell_{\bm{g}_i}$ & $\ell_{\bm{g}_o}$ & $\Rey_\sdist$ & $\ell_\Gamma$ \\[3pt]
          $h^\text{max}_\text{MRI}\simeq2.25h$ & $(\sigma_{u^\star_x},\sigma_{u^\star_y},\sigma_{u^\star_z})$ & $(10,10,10)$ & $10^{-4}$ & $h$ & $h$ & $4$ & $h$\\[6pt]    
  \end{tabular}\\
    \begin{tabular}{cccccc}
           \multicolumn{6}{c}{\emph{numerical parameters (cut-cell FEM)}} \\[3pt]\hline\\[-1.0em]
          $\gamma_\mathcal{N}$ & $\gamma_\nu$  & $\gamma_{u}$ & $\gamma_p$ & $\gamma_{\nabla\text{div}}$ & $c_u$\\[3pt]
          $100$ & $0.05$ & $0.05$ & $0.05$ & $1$ & $1/6$\\[6pt]    
  \end{tabular}\\
  \caption{Problem setup and input parameters for the Bayesian inverse N--S problem.}
  \label{tab:input_params_aorta_3d}
  \end{center}
\end{table}

\subsection{Prior means for the N--S unknowns}
We obtain a level-set function, $\psi$, of the prior mean geometry, $\mean{\Omega}$, using Chan--Vese segmentation \cite{Chan2001,Getreuer2012} on the magnitude images (spin density) of the flow-MRI data (see figure \ref{fig:spin_density_scan}). We then solve the vSDF problem \eqref{eq:vsdf_operator} to obtain the prior mean of the vSDF, $\mean{\varphi}_\pm$. \REV{To regularise the shape of the inferred boundary $\Gamma$ and avoid oversmoothing, we choose a $\Rey_\sdist$ value of 4. Since the smallest scale of geometric details is controlled by the mesh, i.e. $\ell_\Gamma = h$, the diffusion width, $\delta$, of geometric details at distance $h$ from $\Gamma$ is $h/2$ (see section \ref{subsec:geom}).}
The prior mean of the inlet b.c., $\mean{\bm{g}}_i$, is obtained by solving
\begin{gather}
\big(\mathrm{I} - \ell_0^2\widetilde{\Delta}\big)\,\mean{\bm{g}}_i = \big(\mathcal{S}^*\bm{u}^\star\big)\big|_{\Gamma_i} \quad \text{on} \quad \Gamma_i\quad,
\end{gather}
where $\ell_0 = 3h$, $\widetilde{\Delta}$ is the ${L}^2$-extension of the Laplacian that incorporates the zero-Dirichlet (no-slip) b.c. on $\partial\Gamma_i$, and $\big(\mathcal{S}^*\bm{u}^\star\big)\big|_{\Gamma_i}$ is the restriction of the model-space projection of the flow-MRI data on the inlet, $\Gamma_i$. The prior mean of the outlet (natural) b.c., $\mean{\bm{g}}_0$, is set to $\mean{\bm{g}}_0 \equiv (0,0,0)$ due to lack of information, hence the high uncertainty (see table \ref{tab:input_params_aorta_3d}). \REV{The characteristic lengths of the covariance operators for the boundary conditions, $\ell_{\bm{g}_i}$, $\ell_{\bm{g}_o}$ are set to $h$ for regularisation and numerical stability purposes.} Because we used a Newtonian fluid (water) in the flow-MRI experiments, the effective viscosity, $\nu_e$, is assumed to be constant in $\Omega$, and is set to $\mean{\nu}_e = UL/\Rey$. The prior covariance operator, $\priorcov$, is constructed using the values of table \ref{tab:input_params_aorta_3d}, as explained in section \ref{sec:prior_cov_construction}. 

It is important to note that the prior means for $\sdist$ and $\bm{g}_i$ are generated from the flow-MRI data that we reconstruct. That is, we do not perform any additional experiment in order to obtain higher-quality priors (e.g. a computed tomography scan for the geometry or a high-resolution flow-MRI scan for the geometry and the inlet b.c.). The prior means for $\bm{g}_o$ and $\nu_e$, however, cannot be extracted from the flow-MRI scans, and, therefore, depend on the prior knowledge of the problem. 

\subsection{Reconstruction results}
We demonstrate algorithm \ref{algo:bayesian_inversion_ns} at two different flow conditions and two different SNRs for the 3D steady laminar flow in the ROI of the aortic arch (see table \ref{tab:rec_notation} for notation). We apply the algorithm on low SNR flow-MRI data, $\bm{u}^\star$, in order to infer the maximum \emph{a posteriori} (MAP) estimate, $\bm{u}^\circ$, which is the most probable reconstructed velocity field. This highlights the capability of algorithm \ref{algo:bayesian_inversion_ns} to accurately reconstruct flow features that are obscured by noise, which relies on the \emph{explanatory power} of the generalised Navier--Stokes problem. We also apply the algorithm on high SNR flow-MRI data, $\bm{u}^\bullet$, (i.e. $\bm{u}^\star\mapsfrom\bm{u}^\bullet$ in algorithm \ref{algo:bayesian_inversion_ns}) in order to infer the MAP estimate, $\bm{u}^{\bulletcirc}$. This highlights the capability of algorithm \ref{algo:bayesian_inversion_ns} to reproduce (fit) the high-quality velocity images, and helps us identify data-model discrepancies other than Gaussian white noise (i.e. modelling and/or experimental biases).
\begin{table}
\small
\centering
  \begin{tabular}{ccc}
            \emph{data} & \emph{prior} & \emph{reconstruction}\\ \hline\\[-1.0em]
          $\bm{u}^\star/\bm{u}^\bullet$ & $\mean{\bm{u}}/\widetilde{\bm{u}}$ & $\bm{u}^\circ$/$\bm{u}^{\bulletcirc}$
  \end{tabular}\\
\caption{Notation for the low$/$high SNR reconstruction cases.}
\label{tab:rec_notation}
\end{table}

\subsubsection{Velocity field reconstruction at low $\Rey$}
\begin{figure}
\centering
\includegraphics[width=.95\textwidth]{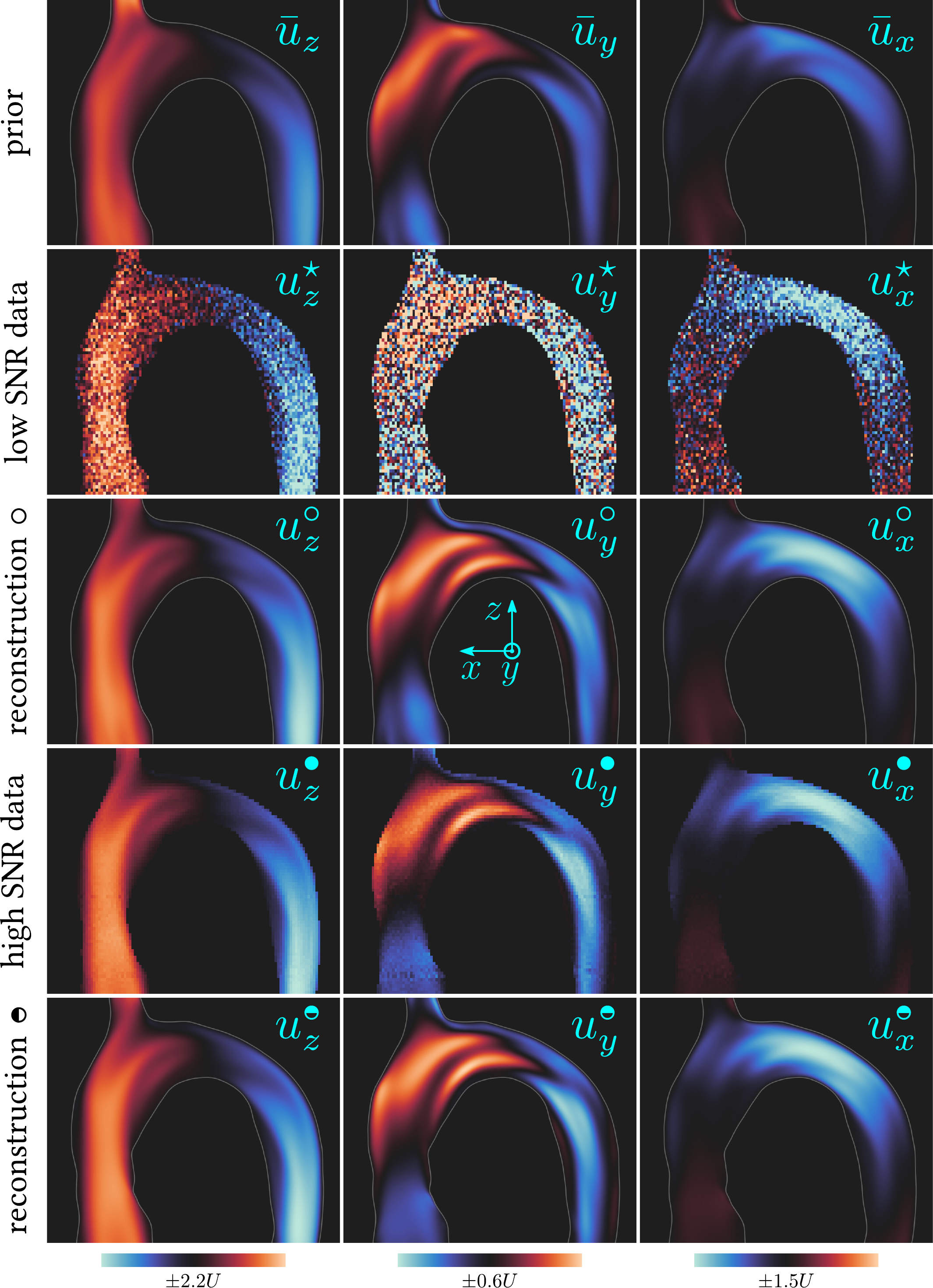}
\caption{We apply algorithm \ref{algo:bayesian_inversion_ns} on low/high SNR data $\bm{u}^\star$/$\bm{u}^\bullet$ for the low $\Rey$ flow ($U = 4.26$ cm/s). The algorithm first generates a prior velocity field, $\mean{\bm{u}}/\widetilde{\bm{u}}$, and then solves a Bayesian inverse N--S problem in order to find the most likely (MAP) reconstruction, $\bm{u}^\circ$/$\bm{u}^{\bulletcirc}$. The prior $\widetilde{\bm{u}}$, which is generated from the high SNR data, $\bm{u}^\bullet$, is similar to $\mean{\bm{u}}$ and thus not shown.}
\label{fig:low_re_rec}
\end{figure}

\begin{figure}
\centering
\begin{subfigure}{\textwidth}
\centering
\includegraphics[width=0.95\textwidth]{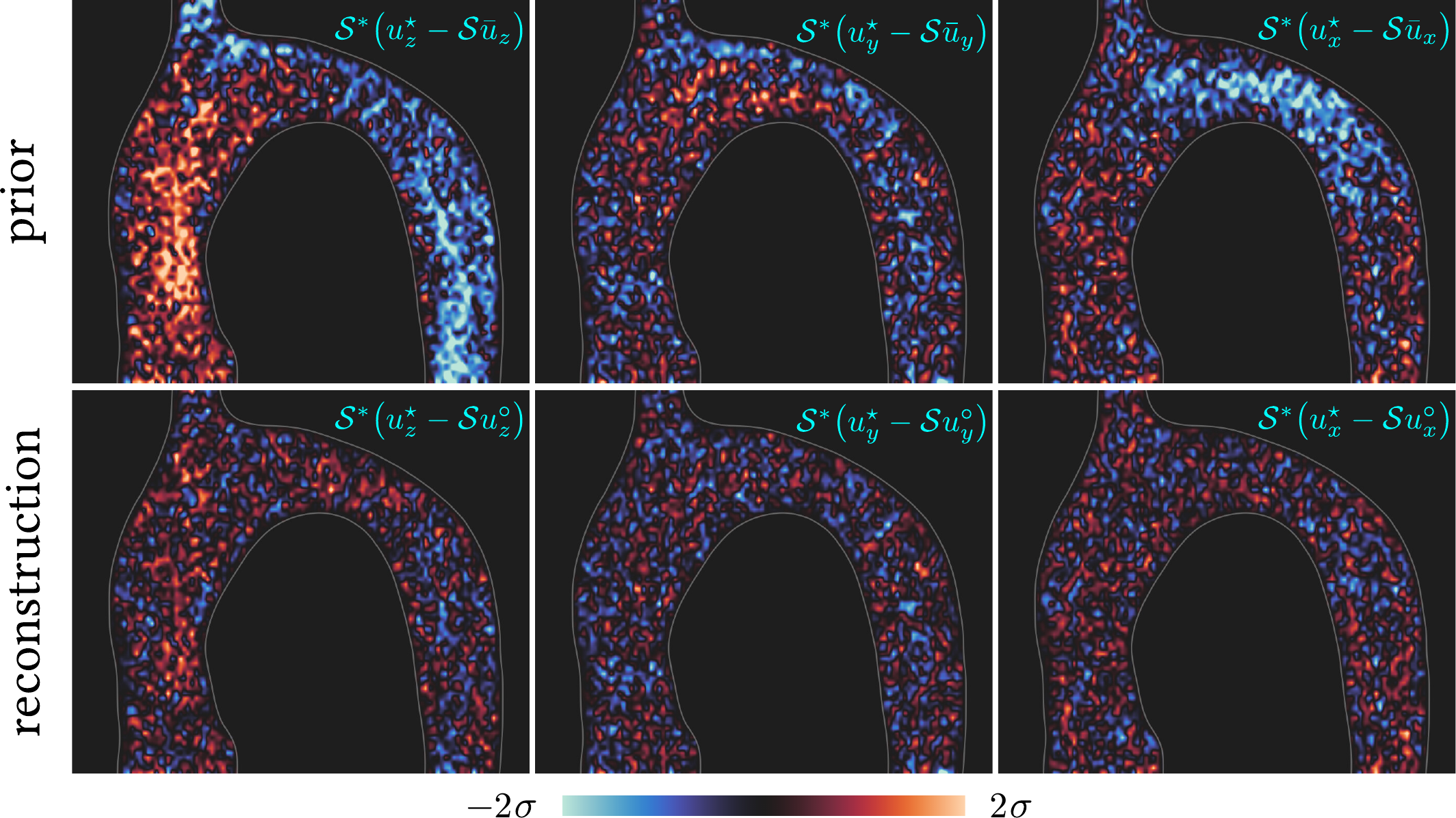}
\caption{velocity residuals at low SNR}
\label{fig:low_re_discrep_lowSNR}
\end{subfigure}\vspace{.25cm}
\begin{subfigure}{\textwidth}
\centering
\includegraphics[width=0.95\textwidth]{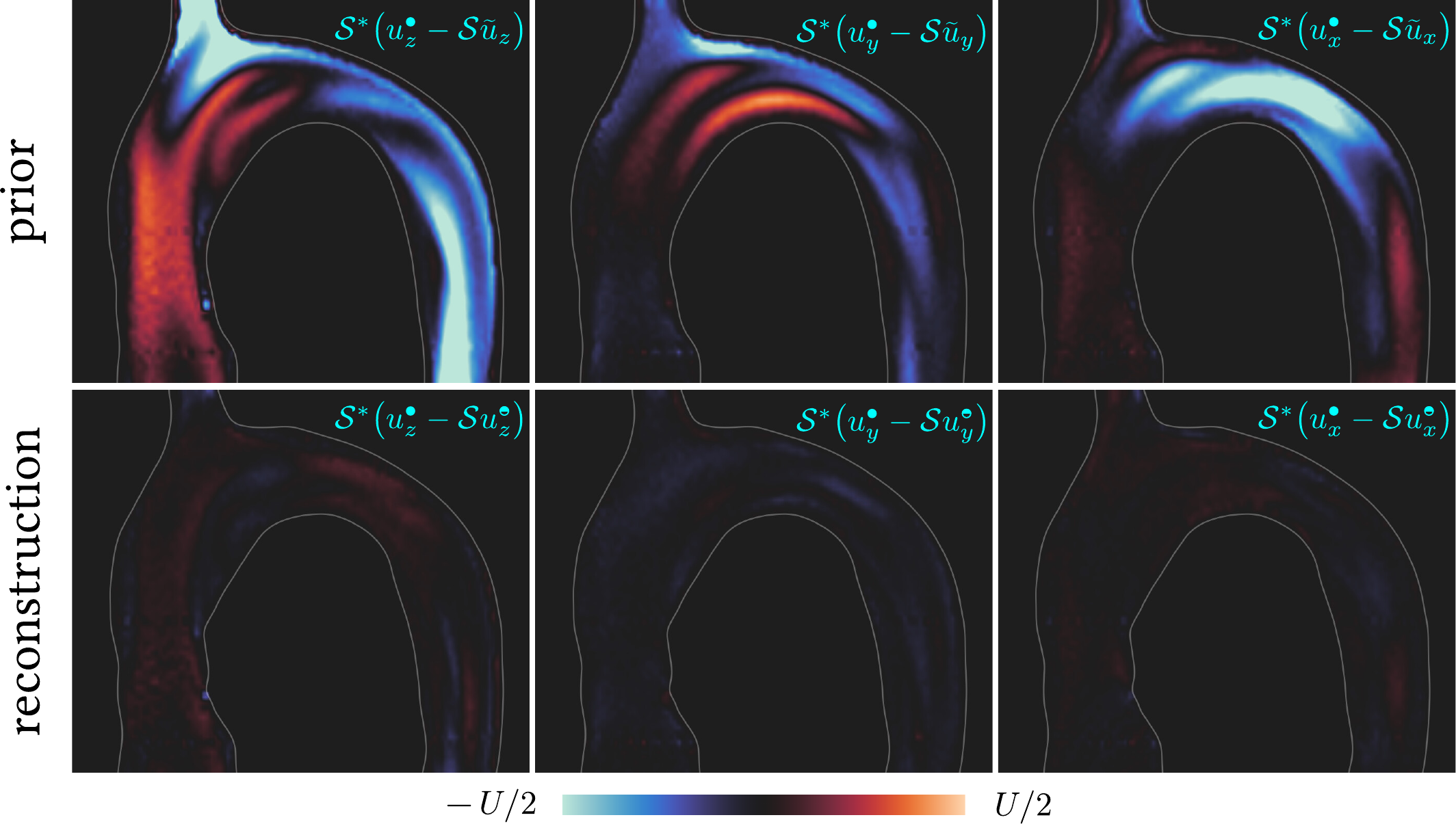}
\caption{velocity residuals at high SNR}
\label{fig:low_re_discrep_highSNR}
\end{subfigure}
\caption{Data-model discrepancies (reconstruction residuals) projected onto the model space, $\bm{M}$, for the low SNR case (figure \ref{fig:low_re_discrep_lowSNR}) and the high SNR case (figure \ref{fig:low_re_discrep_highSNR}) of the low $\Rey$ flow. In the low SNR case, coherent structures in the initial residual (data-prior discrepancy) are assimilated and the final residual (data-MAP discrepancy) approximates Gaussian white noise with variance $\sigma^2$. In the high SNR case the results are similar with the exception that experimental (flow-MRI) biases dominate over noise and the characteristic scale of the error is based on $U$.}
\label{fig:low_re_discrep}
\end{figure}

The reconstruction results for the low $\Rey$ ($=554$) case are shown in figure \ref{fig:low_re_rec}. We observe that the low SNR reconstruction, $\bm{u}^\circ$, obtained from the low SNR data, $\bm{u}^\star$, is almost indistinguishable from the high SNR data, $\bm{u}^\bullet$, and the high SNR reconstruction, $\bm{u}^{\bulletcirc}$, except for minor discrepancies. Also, the high SNR reconstruction, $\bm{u}^{\bulletcirc}$, obtained from the high SNR data, $\bm{u}^\bullet$, fits the data with visually imperceptible error. We reach the same conclusions by inspecting the data-model discrepancies of figure \ref{fig:low_re_discrep}. In the low SNR case, algorithm \ref{algo:bayesian_inversion_ns} manages to assimilate physically relevant information, which appears in the form of coherent structures immersed in Gaussian white noise, while filtering out noise and artefacts. The same applies to the high SNR case, the only difference being that model and/or flow-MRI experimental biases dominate the data-MAP residuals over white noise. For this reason, the characteristic scale for the discrepancy is chosen to be $U$, instead of the noise variance, $\sigma^2$.
For this $\Rey$ number, it is sensible to assume that the bias comes mainly from the flow-MRI experiment because the flow is steady and laminar, and the Newtonian viscosity model describes water to high accuracy. Consequently, the N--S problem \eqref{eq:navierstokes_bvp} can model the velocity field to high accuracy.

\subsubsection{Velocity field reconstruction at high $\Rey$}
\label{sec:rec_results_highRe}
\begin{figure}
\centering
\includegraphics[width=0.95\textwidth]{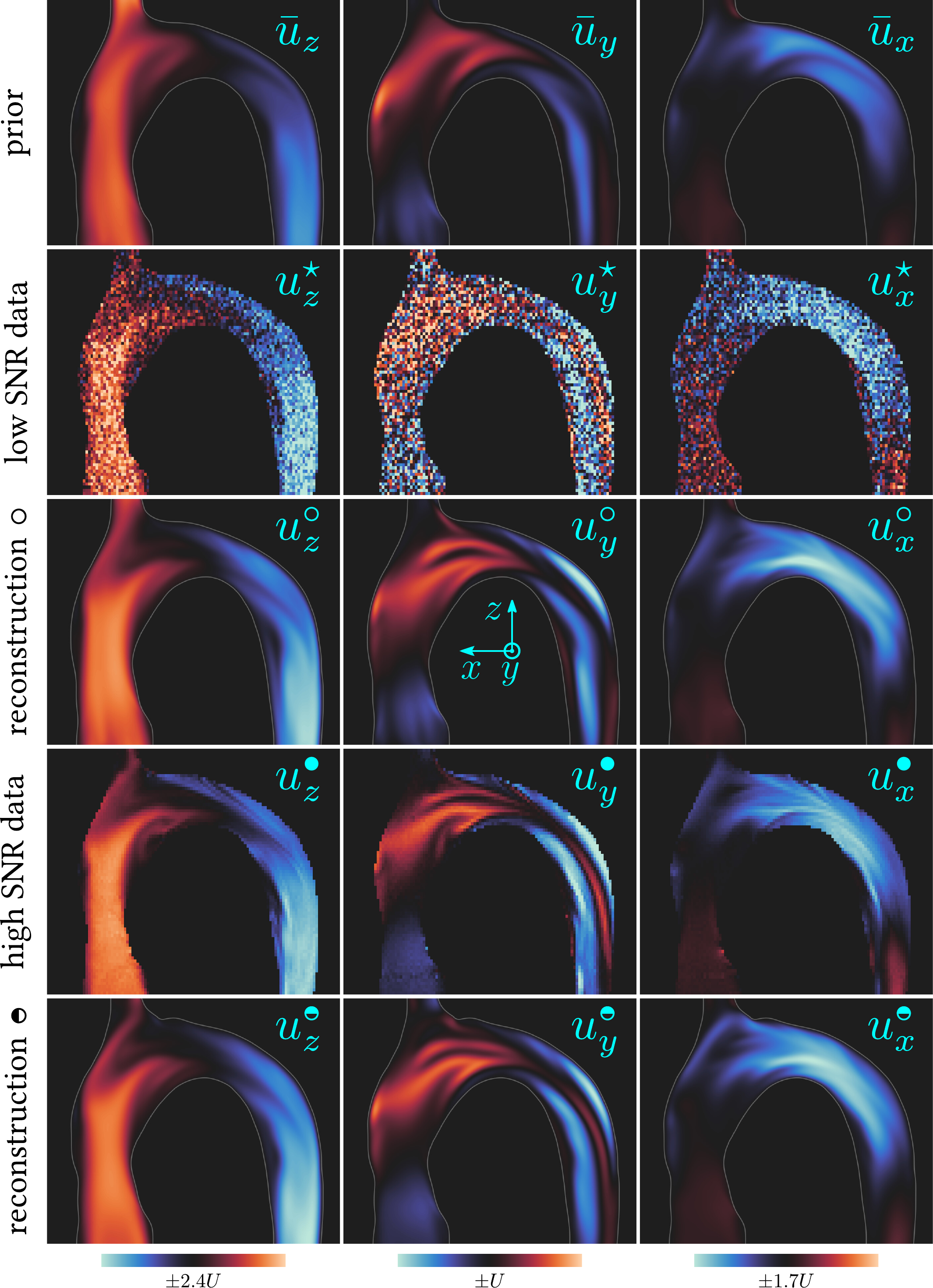}
\caption{As for figure \ref{fig:low_re_rec} but for the high $\Rey$ flow ($U = 11.74$ cm/s).} 
\label{fig:high_re_rec}
\end{figure}

\begin{figure}
\centering
\begin{subfigure}{\textwidth}
\centering
\includegraphics[width=0.95\textwidth]{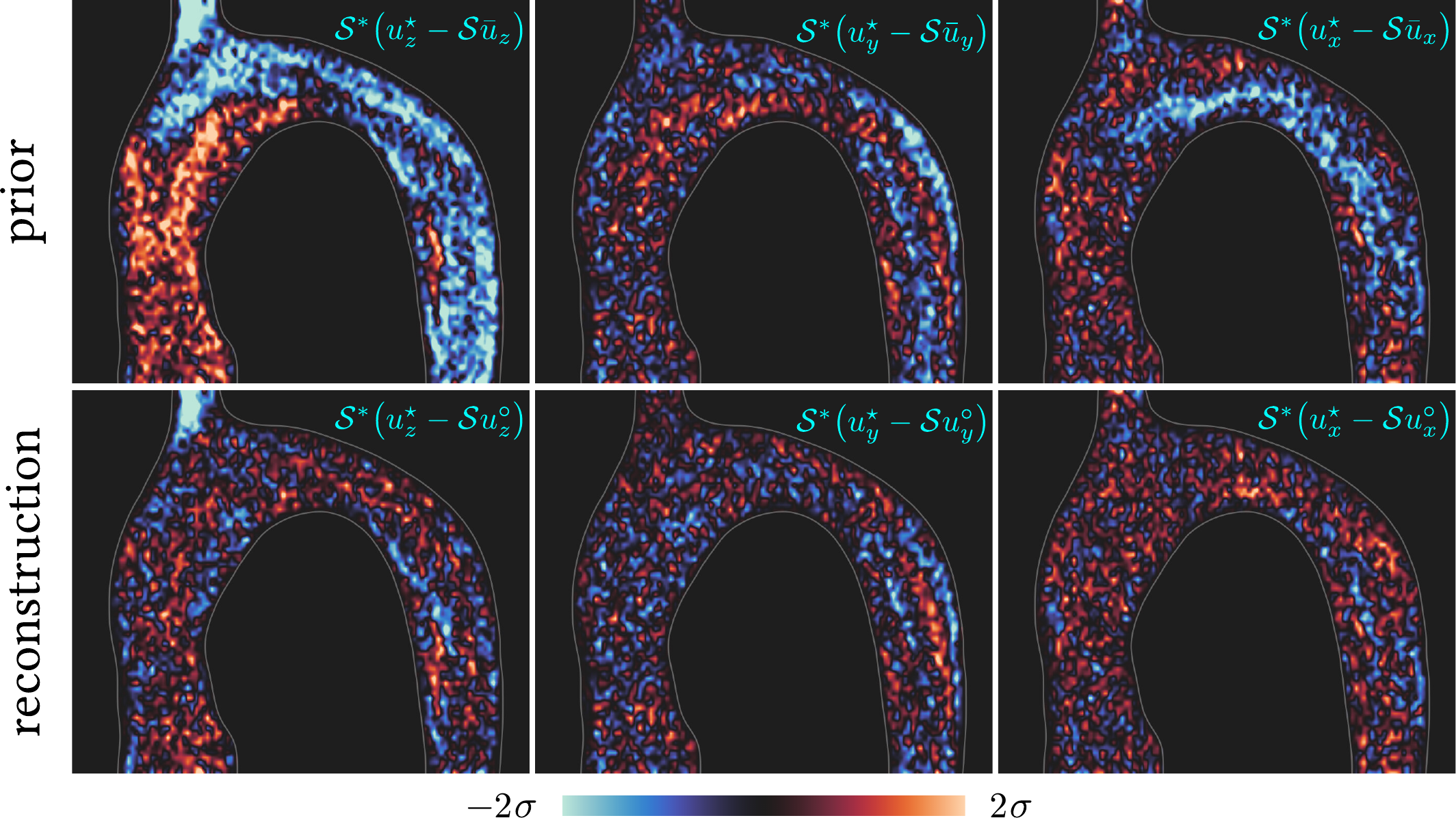}
\caption{velocity residuals at low SNR}
\label{fig:high_re_discrep_lowSNR}
\end{subfigure}\vspace{.25cm}
\begin{subfigure}{\textwidth}
\centering
\includegraphics[width=0.95\textwidth]{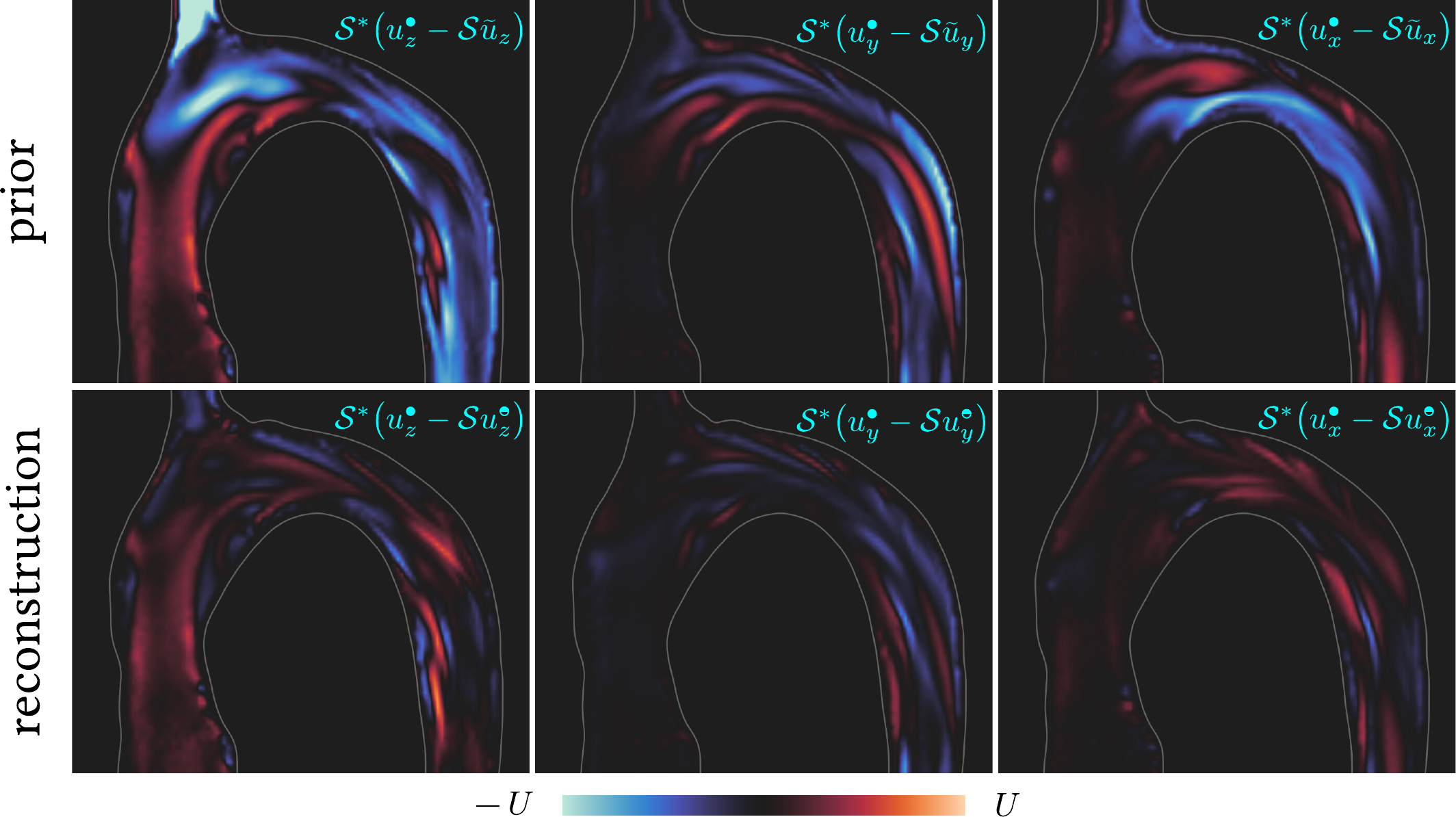}
\caption{velocity residuals at high SNR}
\label{fig:high_re_discrep_highSNR}
\end{subfigure}
\caption{As for figure \ref{fig:low_re_discrep} but for the high $\Rey$ flow.}
\label{fig:high_re_discrep}
\end{figure}

The reconstruction results for the high $\Rey$ ($=1526$) case, shown in figures \ref{fig:high_re_rec} and \ref{fig:high_re_discrep}, can be interpreted in the same way. In the high $\Rey$ case, however, the velocity field has a richer, finer structure, and sharper gradients compared with the low $\Rey$ case. This showcases the capability of algorithm \ref{algo:bayesian_inversion_ns} to reconstruct details of the flow that are completely obscured by noise. In the low SNR case, we observe that the data-MAP residuals (figure \ref{fig:high_re_discrep}) approximate white noise almost everywhere except for two regions: i) the upper main branch, and ii) the descending aorta. Both regions involve recirculation zones and local flow separation (see $\bm{u}^\bullet$ and $\bm{u}^{\bulletcirc}$ in figure \ref{fig:high_re_rec}). In the high SNR case, we observe that the data-MAP residuals are, in contrast to the low $\Rey$ case, non-negligible, and that their peak magnitude is at the recirculation region of the descending aorta. For this near-critical $\Rey$ number, it is possible that the laminar flow transitions to (weakly) turbulent flow in the vicinity of the recirculation bubbles, particularly at the descending aorta\footnote{Qualitative analysis of the MR signal attenuation with and without flow suggests that, at the high $\Rey$ case, these regions of the flow field have weakly fluctuating velocities, which is indicative of locally turbulent flow.} \cite[Figure~5]{MANCHESTER2024108123}. 
Since flow-MRI data are, to a certain extent, averaged (e.g. due to cross-voxel motion and asynchronous spatial and velocity encoding in the MRI sequence), and given that the high SNR flow-MRI scan acquisition time was $\sim3.5$ hours per velocity component \cite[Appendix~E.2]{kontogiannis_2023}, it is possible that flow fluctuations during the course of the experiment (e.g. due to temperature and flow rate changes or bubbles) have slightly distorted the velocity images. It is also expected for this averaging bias to increase in magnitude as the $\Rey$ number increases, since the flow becomes more susceptible to external perturbations.

\subsubsection{Mean reconstruction errors}
\label{sec:mean_rec_error}
We quantitatively compare the reconstructions by defining the mean reconstruction error
\begin{equation}
\bm{\mathcal{E}} \coloneqq \abs{{\Omega}}^{-1}\big\|\bm{u}_{\text{data}} - \mathcal{S}\bm{u}_\text{model}\big\|_{L^2(I_d)}\quad,
\end{equation}
where $\bm{u}_{\text{data}} \in (\bm{u}^\star,\bm{u}^\bullet)$, $\bm{u}_\text{model}\in (\mean{\bm{u}},\widetilde{\bm{u}},\bm{u}^\circ,\bm{u}^{\bulletcirc})$, and $\abs{\Omega}$ is the volume of $\Omega$.

For the low SNR case, we examine the $\sigma_{u^\star}$-normalised errors, $(\mathcal{E}/{\sigma}_{u^\star})_i$, because the noise dominates the residuals (see figures \ref{fig:low_re_discrep_lowSNR}, \ref{fig:high_re_discrep_lowSNR}). Under the assumption of Gaussian white noise, an ideal reconstruction of the low SNR data (i.e. one that perfectly assimilates the data without fitting noise) is expected to have a mean reconstruction error $(\mathcal{E}/{\sigma}_{u^\star})_i = 1$. In practice, however, imperfections in the data, such as artefacts, outliers, and interference noise, increase the kurtosis of the distribution (fat-tailed distribution). Thus, values higher than $1$ may correspond to an ideal reconstruction. In general, under the assumption of Gaussian white noise, $(\mathcal{E}/{\sigma}_{u^\star})_i \gg 1$ implies that the reconstruction algorithm did not succeed in assimilating all the information (e.g. due to experimental or model biases), while $(\mathcal{E}/{\sigma}_{u^\star})_i \ll 1$ implies that the reconstruction overfits the data. In the extreme case of $(\mathcal{E}/{\sigma}_{u^\star})_i = 0$, we have that $\bm{u}_{\text{data}} \equiv \mathcal{S}\bm{u}_\text{model}$, i.e. there is no reconstruction. For the low $\Rey$ case, we find that the MAP reconstruction is ideal, in the sense that the MAP mean errors are near $1.00$ (see table \ref{tab:rec_errors}). For the high $\Rey$ case, the mean errors are larger than $1.00$, which is due to the model/experimental biases that we have attributed to the $\Rey$ number being near-critical (see section \ref{sec:rec_results_highRe}). However, the MAP mean errors are significantly lower than the prior mean errors, showing that the algorithm has managed to assimilate useful information.

For the high SNR case, we examine the errors $(\mathcal{E}/U)_i$ because the experimental/model bias dominates the residuals (see figures \ref{fig:low_re_discrep_highSNR}, \ref{fig:high_re_discrep_highSNR}). We find that for both the low and high $\Rey$ cases there is a significant reduction in the mean error (see table \ref{tab:rec_errors}). For the low $\Rey$ case, the MAP reconstruction fits the high SNR data with high accuracy and the mean errors are negligible. For the high $\Rey$ case, the mean errors are non-negligible due to the model/experimental biases that we have attributed to the near-critical $\Rey$ number (see section \ref{sec:rec_results_highRe}). For comparison, volumetric flow rates calculated from the flow-MRI images, in the ascending portion of the aorta, agreed with volumetric flow rates measured from the pump outlet within an average error of $\pm0.04U$.

\begin{table}[h]
\small
\centering
  \begin{tabular}{ccc}
           \multicolumn{3}{c}{\emph{prior velocity field}} \\[3pt]\hline\\[-1.0em]
          & low SNR ($\bm{\mathcal{E}}/\bm{\sigma}_{u^\star}$) & high SNR ($\bm{\mathcal{E}}/U$) \\[3pt]
          low $\Rey$ & $(1.09,1.04,1.23)$ & $(0.21,0.14,0.30)$ \\[3pt]
          high $\Rey$ & $(1.20,1.21,1.54)$& $(0.32,0.34,0.52)$ \\[8pt]
          \multicolumn{3}{c}{\emph{posterior (MAP) velocity field}} \\[3pt]\hline\\[-1.0em]
          & low SNR ($\bm{\mathcal{E}}/\bm{\sigma}_{u^\star}$) & high SNR ($\bm{\mathcal{E}}/U$) \\[3pt]  
          low $\Rey$ & $(1.00,1.00,1.02)$ & $(0.04,0.04,0.06)$ \\[3pt]
          high $\Rey$ & $(1.07,1.06,1.12)$& $(0.17,0.15,0.24)$ \\[6pt]
  \end{tabular}\\
\caption{Mean reconstruction errors for the prior and the posterior (MAP) velocity fields.}
\label{tab:rec_errors}
\end{table}

\begin{figure}
\centering
\begin{subfigure}{.425\textwidth}
\centering
\includegraphics[width=.95\textwidth]{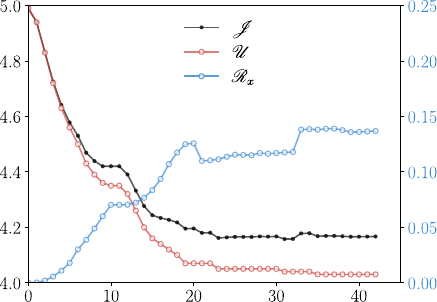}
\caption{low $\Rey$, low SNR}
\end{subfigure}
\begin{subfigure}{.425\textwidth}
\centering
\includegraphics[width=.95\textwidth]{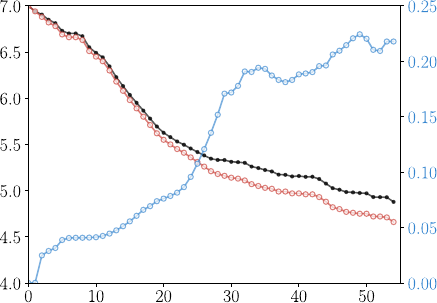}
\caption{high $\Rey$, low SNR}
\end{subfigure}
\begin{subfigure}{.425\textwidth}
\centering
\includegraphics[width=.95\textwidth]{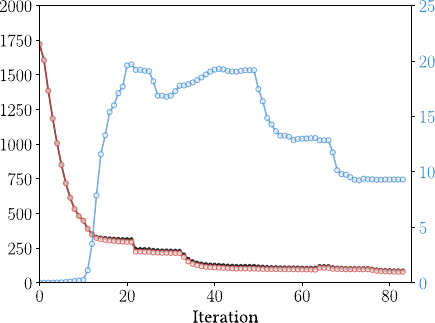}
\caption{low $\Rey$, high SNR}
\end{subfigure}
\begin{subfigure}{.425\textwidth}
\centering
\includegraphics[width=.95\textwidth]{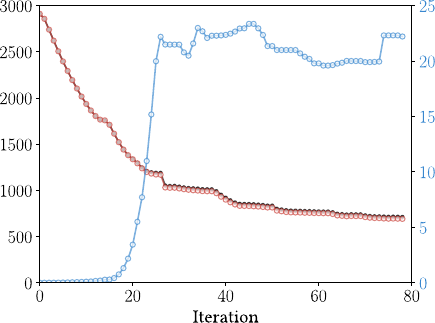}
\caption{high $\Rey$, high SNR}
\end{subfigure}
\caption{Optimisation log depicting the evolution of the objective functional, $\mathscr{J}$, the data-model discrepancy, $\mathscr{U}$, and the priors, $\mathscr{R}_{\bm{x}}$, for the four different inverse problems.}
\label{fig:optimisation_log}
\end{figure}
Figure \ref{fig:optimisation_log} shows the optimisation logs of the four reconstruction problems. Note that the high SNR reconstruction problems converge more slowly than the low SNR reconstruction problems because there is more information to be assimilated. 

\subsection{Explanatory power and interpretability of the N--S problem}
It is worth noting that the N--S unknowns, $\bm{x}$, as chosen in this study, correspond to physical quantities that are \emph{interpretable}. For example, in cardiovascular applications, the fluid domain, $\Omega$, defines the geometry of the blood vessel, the boundary condition $\bm{g}_i$ defines the inlet velocity, the boundary condition $\bm{g}_o$ defines the fluid stresses at the outlets, and the effective kinematic viscosity, $\nu_e$, depends on the rheological properties of the fluid \cite{Kontogiannis2024b}. Although we have neglected the equations of elastodynamics, we have not fixed the shape of $\Omega$. Therefore, for unsteady flows and compliant walls, we can still infer the movement of the geometry and its influence on the velocity field, but we cannot yet infer the elastic properties of the walls. 

Another important aspect regarding the choice of the N--S parameters, $\bm{x}$, is that they have limited control over the modelled velocity field, $\bm{u}$. More precisely, every N--S unknown is a function defined on a \mbox{$(d-1)$-manifold}, where $d$ is the dimension of the problem. This means that the inverse N--S problem is a \emph{sparsifying transform}\footnote{In other words, the inverse N--S problem can be used for data compression and storage of flow-MRI data.} \cite{Kontogiannis2022b} that maps velocity fields, defined on a subset of $\mathbb{R}^d$, to N--S unknowns, defined on a subset of $\mathbb{R}^{d-1}$ (see table \ref{tab:dim_reduction}). In contrast, the inverse, \emph{forced} N--S problem, which is obtained after adding a volumetric forcing term, $\bm{f}$, to the momentum equation in order to enhance control (i.e. increase model flexibility to fit the data), is no longer a sparse transformation. It can be made sparse, however, if $\bm{f}$ is parametrised such that $\bm{f}=\bm{f}(\bm{z})$, where $\bm{z}$ is an unknown defined on a $(d-1)$-manifold or a subset thereof. In addition, the inverse problem is more interpretable if $z$ has a physical meaning, which is preferable. 

Because we limit the control to physically-interpretable parameters in $\mathbb{R}^{d-1}$, we rely on the \emph{explanatory power} of the N--S problem, which relies on first principles, such as momentum and mass conservation, and explains much from little. This an advantage because it provides insight into the physical problem. If the model cannot explain the data then it has to be revised, which leads to an improved physical model and the advancement of fluid dynamics modelling. The process of model selection can be made even more rigorous using Bayesian inference \cite[Chapter~28]{MacKay2003}\cite{JUNIPER2022117096,Yoko_Juniper_2024}.

\begin{table}
\small
\centering
    \begin{tabular}{cccc|cc|c}
           \multicolumn{4}{c}{\emph{unknowns (search space)}} & \multicolumn{2}{c}{\emph{model states}} & voxels\\[3pt]\hline\\[-1.0em]
           $\Gamma$ & $\bm{g}_i$ & $\bm{g}_o$ & $\nu_e$ &$(\bm{u},p)$  & $\sdist$ & $\abs{\mathcal{T}_h}$\\[3pt]
          $\mathcal{O}(n^2)$ & $\mathcal{O}(n^2)$ & $\mathcal{O}(n^2)$ & $\mathcal{O}(1)$ & $\mathcal{O}(n^3)$ & $\mathcal{O}(n^3)$ & $\mathcal{O}(n^3)$\\[6pt]    
  \end{tabular}\\
  \caption{Dimensionality of the problem assuming a discrete domain of $\mathcal{O}(n^3)$ voxels.}
  \label{tab:dim_reduction}
\end{table}

\section{Summary and conclusions}
We have formulated a Bayesian inverse N--S problem that assimilates noisy and sparse velocimetry data in order to jointly reconstruct the flow field and infer the unknown N--S parameters. By hardwiring a generalised N--S problem and regularising its unknown parameters using Gaussian prior distributions, the method learns the most likely (MAP) parameters in a collapsed search space. The most likely (MAP) velocity field reconstruction is then the N--S solution that corresponds to the MAP parameters. Working in a variational setting, we use a stabilised Nitsche weak form of the N--S problem that permits control of all N--S parameters, and, in particular, the Dirichlet (inlet velocity) boundary condition. We implicitly define the geometry using a viscous signed distance field, and control its regularity by choosing an appropriate artificial diffusion coefficient based on (prior) boundary smoothness assumptions. The vSDF is also implicitly defined as the solution of a viscous Eikonal boundary value problem, which is incorporated as a second constraint (the first being the N--S problem) in the saddle point problem. To solve the Bayesian inverse N--S problem we devise an iterative algorithm that searches for \emph{saddle points} of the objective functional by attempting to satisfy the first order optimality conditions (i.e. solves the Euler--Lagrange system). We approximate the inverse Hessian operator of the N--S parameters, which is equivalent to the posterior covariance parameter operator, using a quasi-Newton method (BFGS). This allows us to: i) accelerate the iterative procedure using curvature information (gradient preconditioning), and ii) estimate parameter uncertainties around the MAP point (Laplace approximation). We numerically implement the algorithm using a fictitious domain cut-cell finite element method, and stabilise the numerical problem by adding symmetric and consistent stabilisation terms ($\nabla$-div and CIP) to the continuous weak form. Consequently, the numerical solution converges to the continuous solution as the background mesh size tends to zero, and the discrete formulation can be shown to be adjoint-consistent (i.e. the discretised continuous adjoint operator, $(\mathcal{G}^*)_h$, is equivalent to the discrete adjoint operator, $(\mathcal{G}_h)^*$).

We then apply the method to the reconstruction of flow-MRI data of a steady laminar flow through a 3D-printed physical model of an aortic arch for two different Reynolds numbers and SNR levels (low/high). We find that the method can accurately i) reconstruct the low SNR data by filtering out the noise/artefacts and recover flow features that are completely obscured by noise, and ii) reproduce (fit) the high SNR data using limited control and without overfitting (e.g. there is no volumetric forcing term in the N--S problem). In the low $\Rey$ case, the reconstructed (MAP) low SNR data are indistinguishable from the high SNR data, and the mean error of the reconstructed (MAP) high SNR data is approximately equal to the mean error of the flow-MRI experiment (as seen through the volumetric flow rate comparison between flow-MRI data and pump outlet). In the high $\Rey$ case the results are similar, but the mean reconstruction errors, although small, are non-negligible due to model/experimental biases (i.e. \mbox{flow-MRI} averaging and, possibly, locally turbulent flow) that we have attributed to the $\Rey$ number being near-critical. 

The method described herein \REV{could} increase the accuracy and the spatiotemporal resolution
of velocity imaging (e.g. flow-MRI, PIV, and Doppler velocimetry) and/or significantly decrease scanning times, thus enabling the imaging of flows whose short length and/or time scales cannot be captured using state-of-the-art imaging techniques (e.g. smaller vessels such as those found in neonatal and fetal cardiology). By jointly reconstructing the flow field and learning the unknown N--S parameters, as well as their uncertainties, the method provides quantitative estimates of other derived flow quantities that are difficult to measure (e.g. pressure, wall shear stress and rheological parameters). The method requires no training, can explain much from little, is extrapolatable and is interpretable, because it hardwires prior fluid dynamics knowledge in the form of a generalised N--S problem. Consequently, it learns the most likely digital twin of the experiment, which has interesting applications in engineering design and patient-specific cardiovascular modelling. Lastly, by appropriately choosing the N--S unknowns, the inverse N--S problem is shown to be a sparsifying transform of the 3D (or 4D) velocity field. Since only the most likely (MAP) N--S unknowns need be stored to generate the flow field reconstruction, this allows the simultaneous reconstruction and compression of velocimetry data.

\subsection{Future work: extensions and other applications}
In this paper, we have limited our study to 3D steady laminar flows. The framework that we develop, however, readily extends to time-dependent laminar and turbulent flows, be they periodic or unsteady flows, and compliant walls. In particular, the formulation can be extended to reconstruct Reynolds-averaged turbulent flows and non-Newtonian fluid flows. This has already been done for power law fluids \cite{Kontogiannis2024b}, and it further extends to viscoelastic fluids. Concerning the numerics, i.e. the discrete formulation, the algorithm can greatly benefit from adaptive discretisation and multiresolution methods (in this study we used a uniform Cartesian mesh, which is suboptimal and computationally expensive). Even though the algorithm in this paper reconstructs velocimetry data, it can be extended to reconstruct raw, sparse, and noisy flow-MRI data in frequency space, as in \cite{Kontogiannis2022b}. Lastly, the Bayesian inversion framework that we adopt is already set up for model comparison (e.g. ranking different viscoelastic or turbulence models based on their evidence), experimental/model bias learning (i.e. inferring the unknown unknowns) \cite[Section~4.2.3]{Yoko_Juniper_2024}, and optimal experiment design \cite{Yoko_Juniper_2024b} (e.g. design of optimal $\bm{k}$-space sampling patterns in order to accelerate flow-MRI scanning for specific types of flows).

\subsubsection*{Author contributions}
\textbf{AK.} conceptualization, formal analysis, funding acquisition, investigation, methodology, project administration, software, validation, visualization, writing -- original draft, writing -- review \& editing. \textbf{SVE.} data curation, investigation, writing -- review \& editing. \textbf{AJS.} funding acquisition, project administration, resources, supervision, writing -- review \& editing. \textbf{MPJ.} conceptualization, funding acquisition, project administration, resources, supervision, writing -- review \& editing.


\subsubsection*{Author ORCIDs}
\href{https://orcid.org/0000-0001-6353-3427}{\includegraphics[scale=0.05]{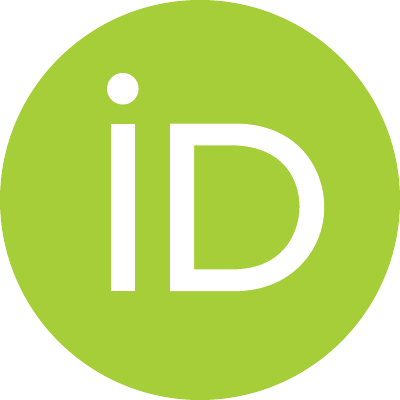}} Alexandros Kontogiannis \href{https://orcid.org/0000-0001-6353-3427}{https://orcid.org/0000-0001-6353-3427};\\
\href{https://orcid.org/0000-0001-6353-3427}{\includegraphics[scale=0.05]{Figs/ORCID_iD}} Scott V. Elgersma \href{https://orcid.org/0000-0001-6353-3427}{https://orcid.org/0009-0002-4179-9747};\\
\href{https://orcid.org/0000-0001-6353-3427}{\includegraphics[scale=0.05]{Figs/ORCID_iD}} Andrew J. Sederman \href{https://orcid.org/0000-0001-6353-3427}{https://orcid.org/0000-0002-7866-5550};\\
\href{https://orcid.org/0000-0001-6353-3427}{\includegraphics[scale=0.05]{Figs/ORCID_iD}} Matthew P. Juniper \href{https://orcid.org/0000-0001-6353-3427}{https://orcid.org/0000-0002-8742-9541}.

\subsubsection*{Acknowledgement}
AK is supported by EPSRC National Fellowships in Fluid Dynamics (NFFDy), grant EP/X028232/1.

\subsubsection*{Data availability statement}
The experimental low SNR flow-MRI data used in this study are available at \url{https://doi.org/10.17863/CAM.109745}.

\begin{appendices}

\section{Trace operators}
\label{app:ext_trace_ops}
In order to smoothly extend the boundary conditions $\bm{g}_i$, $\bm{g}_o$ to the whole domain, $I_m$, we introduce a modified adjoint of the trace operator. We define this adjoint trace operator, $T^*_{A}$, such that $\bm{v} \equiv T^*_A~\bm{g}$ is equivalent to solving the boundary value problem
\begin{gather}
\big(\bm{\nuext\bm{\cdot}}\nabla - \eps \Delta\big)\bm{v} = \bm{0} \quad \text{in}~\Omega\quad,\quad \bm{v}=\bm{g}\quad\text{on}~A\quad,
\end{gather}
where $\nuext$ is outward-facing unit normal vector extension, which is given by \eqref{eq:normal_vec_extension}, and $\R^+\ni\eps \ll 1$ (in this study we take $\eps=h$, where $h$ is the mesh size).

\section{Leibniz--Reynolds transport theorem}
\label{app:transp_thm}
To find the shape derivative of an integral defined in $\Omega$, when the boundary $\partial\Omega$ deforms with speed $\spd$, we use the Leibniz--Reynolds transport theorem. For the bulk integral of $f:\Omega \to \R$, we find
\begin{gather}
\frac{d}{d\tau}\bigg(\int_{\Omega(\tau)} f\bigg)\bigg\vert_{\tau=0} = \int_\Omega f' + \int_{\partial\Omega}f~(\spd\bm{\cdot}\bm{\nu})\quad,
\label{eq:reynolds_bulk}
\end{gather} 
while for the boundary integral of $f$ we find \cite[Chapter~5.6]{Walker2015}
\begin{gather}
\frac{d}{d\tau}\bigg(\int_{\partial\Omega(\tau)} f\bigg)\bigg\vert_{\tau=0} = \int_{\partial\Omega} f' + (\partial_{\bm{\nu}}+\kappa)f ~(\spd\bm{\cdot}\bm{\nu})\quad,
\label{eq:reynolds_bound}
\end{gather}
where $f'$ is the shape derivative of $f$ (due to $\spd$), $\kappa$ is the summed curvature of $\partial\Omega$, and $\spd \equiv \zeta\bm{\nu}$, with $\zeta \in L^2(\partial\Omega)$, is the Hadamard parameterisation of the speed field. Any boundary that is a subset of $\partial I_m$, i.e. the boundary of the model space $I_m$, is non-deforming and therefore the second term of the above integrals vanishes. The only boundary that deforms is $\Gamma$ ($\Gamma\subset \partial\Omega \subset I_m$). 
\end{appendices}

\printbibliography

\end{document}